\documentclass[aps,amsfonts,12pt]{article}

\usepackage{color}
\usepackage{amsmath,amssymb}
\usepackage{epsfig}
\usepackage{subfigure}
\usepackage{axodraw4j}
\usepackage{a4wide}
\usepackage{multicol}

\usepackage{cite} 

\usepackage{graphicx}
\usepackage{amsmath}
\usepackage{amsbsy}

\newcommand{\ee}{\end{equation}}
\newcommand{\bb}{\begin{equation}}
\newcommand{\eqb}{\begin{eqnarray}}
\newcommand{\eqf}{\end{eqnarray}}

\definecolor{purple}{rgb}{0.4,0,0.6}
\definecolor{gre}{rgb}{0,0.7,0.3}

\newcommand{\pb}{{ \beta}}

\begin{document}

\title{\vspace{-3cm}{\small \hfill{{DESY 10-147; IPPP/10/77;
DCPT/10/154; MPP-2010-123}}}\\[1.8cm]
Optimizing Light-Shining-through-a-Wall Experiments for Axion and other WISP Searches}
\author{Paola Arias$^{a,}$\footnote{{\bf
e-mail}: paola.arias@desy.de} \,\,, Joerg Jaeckel$^{b,}$\footnote{{\bf
e-mail}: joerg.jaeckel@durham.ac.uk}\,\,, Javier Redondo$^{c,}$\footnote{{\bf
e-mail}: redondo@mppmu.mpg.de}\,\,, and Andreas Ringwald$^{a,}$\footnote{{\bf
e-mail}: andreas.ringwald@desy.de}\,\,
\\[2ex]
\small{\em $^a$Deutsches Elektronen-Synchrotron, Notkestra\ss e 85, D-22607 Hamburg, Germany}\\
\small{\em $^b$Institute for Particle Physics Phenomenology, Durham University, Durham DH1 3LE, UK}\\
\small{\em $^c$Max-Planck-Institute f\"ur Physik, F\"ohringer Ring 6, D-80805 M\"unchen, Germany}}

\date{}

\maketitle

\begin{abstract}
One of the prime tools to search for new light bosons interacting very weakly with photons -- prominent examples are
axions, axion-like particles and extra ``hidden'' U(1) gauge bosons -- are light-shining-through-a-wall (LSW)
experiments. With the current generation of these experiments finishing data taking it is time to plan for the next and
search for an optimal setup. The main challenges are clear: on the one hand we want to improve the sensitivity towards
smaller couplings, on the other hand we also want to increase the mass range to which the experiments are sensitive. Our
main example are axion(-like particle)s but we also discuss implications for other WISPs (weakly interacting slim
particles) such as hidden U(1) gauge bosons. To improve the sensitivity for axions towards smaller couplings one can use
multiple magnets to increase the length of the interaction region. However, naively the price to pay is that the mass
range is limited to smaller masses. We discuss how one can optimize the arrangement of magnets (both in field direction
as well as allowing for possible gaps in between) to ameliorate this problem. Moreover, future experiments will include
resonant, high quality optical cavities in both the production and the regeneration region. To achieve the necessary high
quality of the cavities we need to avoid too high diffraction losses. This leads to minimum requirements on the diameter
of the laser beam and therefore on the aperture of the cavity. We investigate what can be achieved with currently
available magnets and desirable features for future ones.

\end{abstract}

\section{Introduction}
Many extensions of the Standard Model (SM) predict a ``hidden sector" of
particles which transform trivially under the SM gauge group and therefore interact only very weakly with the
``visible" sector, {\it{i.e.}} the particles of the SM. Due to their feeble interactions such particles are notoriously difficult to detect and
even very light ones, with masses below an eV, may have escaped detection so far.
Such new very weakly interacting sub-eV particles have been dubbed WISPs (weakly interacting slim particles).
Beyond this purely phenomenological reasoning we have good theoretical motivation for the existence of light but also
very weakly interacting particles. The prime example is the axion which arises as a very light pseudo-Goldstone boson in the
course of a solution of the strong CP problem~\cite{Peccei:1977hh,Weinberg:1977ma,Wilczek:1977pj,Kim:1979if,Dine:1981rt,Shifman:1979if,Zhitnitsky:1980tq}.
Here, because of the pseudo-Goldstone nature,
the weakness of the couplings $\sim 1/f_{a}$
and the smallness of the mass\footnote{$m_{\pi},f_{\pi}$ are the known pion mass and decay constant.}
$m_{a}\sim m_{\pi}f_{\pi}/f_{a}\sim {\rm meV}(10^{10} {\rm GeV}/f_{a})$
are inherently related to a high energy scale $f_{a}$ at which the breaking of the Peccei-Quinn symmetry occurs.
So, not only it is very plausible to have light, very weakly coupled particles, but indeed if we find them we may obtain information
on underlying physics at very high energy scales.

Similarly other light (pseudo-)scalar bosons with analogous couplings but possibly different masses, so called axion-like
particles (ALPs), arise quite naturally. Indeed top-down models such as those arising from compactifications in string
theory suggest plenty of ALPs~\cite{Witten:1984dg,Conlon:2006tq,Svrcek:2006yi,Arvanitaki:2009fg}. Other particularly well motivated WISPs
candidates are hidden U(1) gauge bosons \cite{Okun:1982xi,Holdom:1985ag} and mini-charged particles \cite{Holdom:1985ag}.
They naturally arise in models based on supergravity or
superstrings~\cite{Dienes:1996zr,Abel:2003ue,Abel:2006qt,Abel:2008ai,Goodsell:2009xc,Goodsell:2010ie}.
All in all, the detection (or also the
non-detection) of WISPs may give us important information on the underlying structure of fundamental physics.
Although we keep an eye on general WISPs, the main focus of this paper will be the optimization of experiments searching for ALPs.

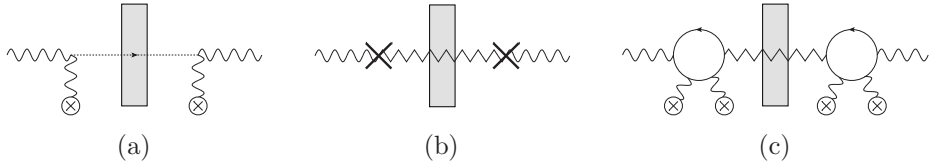
\begin{figure}[t!]
\begin{center}
\subfigure[]{\scalebox{0.3}[0.3]{
  \begin{picture}(322,141) (95,-63)
    \SetWidth{1.0}
    \SetColor{Black}
    \Photon(96,13)(176,13){7.5}{4}
    \GBox(240,-51)(272,77){0.882}
    \Line[dash,dashsize=2,arrow,arrowpos=0.5,arrowlength=5,arrowwidth=2,arrowinset=0.2](176,13)(336,13)
    \Photon(176,13)(176,-51){7.5}{3}
    \COval(176,-51)(11.314,11.314)(45.0){Black}{White}\Line(170.343,-56.657)(181.657,-45.343)\Line(170.343,-45.343)(181.657,-56.657)
    \Photon(336,13)(336,-51){7.5}{3}
    \Photon(336,13)(416,13){7.5}{4}
    \COval(336,-51)(11.314,11.314)(45.0){Black}{White}\Line(330.343,-56.657)(341.657,-45.343)\Line(330.343,-45.343)(341.657,-56.657)
  \end{picture}
  }
\label{fig:lswa}}
  \hspace*{0.2cm}
\subfigure[]{\scalebox{0.3}[0.3]{
  \begin{picture}(322,130) (95,-73)
    \SetWidth{1.0}
    \SetColor{Black}
    \Photon(96,2)(176,2){7.5}{4}
    \GBox(240,-62)(272,66){0.882}
    \Photon(336,2)(416,2){7.5}{4}
    \ZigZag(176,2)(336,2){7.5}{8}
    \SetWidth{3.0}
    \Line(160.002,17.998)(191.998,-13.998)\Line(191.998,17.998)(160.002,-13.998)
    \Line(320.002,17.998)(351.998,-13.998)\Line(351.998,17.998)(320.002,-13.998)
  \end{picture}
  }
\label{fig:lswb} } \hspace*{0.2cm}
 \subfigure[]{\scalebox{0.3}[0.3]{
  \begin{picture}(386,141) (63,-63)
    \SetWidth{1.0}
    \SetColor{Black}
    \GBox(240,-51)(272,77){0.882}
    \Photon(128,13)(64,13){7.5}{3}
    \Arc[arrow,arrowpos=0.5,arrowlength=5,arrowwidth=2,arrowinset=0.2](160,13)(32,270,630)
    \ZigZag(192,13)(320,13){7.5}{6}
    \Arc[arrow,arrowpos=0.5,arrowlength=5,arrowwidth=2,arrowinset=0.2](352,13)(32,270,630)
    \Photon(384,13)(448,13){7.5}{3}
    \Photon(146,-16)(127,-52){7.5}{2}
    \Photon(176,-15)(192,-51){7.5}{2}
    \Photon(336,-15)(320,-51){7.5}{2}
    \Photon(368,-14)(384,-52){7.5}{2}
    \COval(128,-51)(11.314,11.314)(45.0){Black}{White}\Line(122.343,-56.657)(133.657,-45.343)\Line(122.343,-45.343)(133.657,-56.657)
    \COval(192,-51)(11.314,11.314)(45.0){Black}{White}\Line(186.343,-56.657)(197.657,-45.343)\Line(186.343,-45.343)(197.657,-56.657)
    \COval(320,-51)(11.314,11.314)(45.0){Black}{White}\Line(314.343,-56.657)(325.657,-45.343)\Line(314.343,-45.343)(325.657,-56.657)
    \COval(384,-51)(11.314,11.314)(45.0){Black}{White}\Line(378.343,-56.657)(389.657,-45.343)\Line(378.343,-45.343)(389.657,-56.657)
  \end{picture}
  }
\label{fig:lswc}}
  \end{center}
\caption{\footnotesize{Basic principle of an LSW experiment. An incoming photon $\gamma$ is converted into a new particle
which interacts only very weakly with the opaque wall. It passes through the wall and is subsequently reconverted into an
ordinary photon which can be detected. The different panels show the explicit processes contributing to LSW for various
WISPs. From left to right we have photon -- ALP, photon -- hidden photon and photon -- hidden photon oscillations
facilitated by MCPs.}}\label{fig:lsw}
\end{figure}

Optical precision experiments are a powerful tool to search for WISPs, thereby exploring the hidden sector. In
particular, laser experiments have an enormous potential to search for particles with tiny couplings to photons. A class
of very simple and effective laser experiments is based on photon -- WISP -- photon oscillations: the so called {\it light
shining through a wall} (LSW) experiments. Let us briefly recap the basic principle as shown in Fig.~\ref{fig:lswa} for
the case of ALPs (Figs.~\ref{fig:lswb},\ref{fig:lswc} show the relevant processes for other WISPs). If light is shone
through a magnetic region, a small fraction of the beam can be converted into ALPs, exploiting the Primakoff
effect~\cite{Sikivie:1983ip,Anselm:1987vj,VanBibber:1987rq}. A barrier after the magnetic regions allows only the ALP component
of the beam to enter into a second identical magnetic region, where the ALPs can be reconverted into photons.

\begin{figure}[t]
\begin{center}
\includegraphics[width=0.85\textwidth]{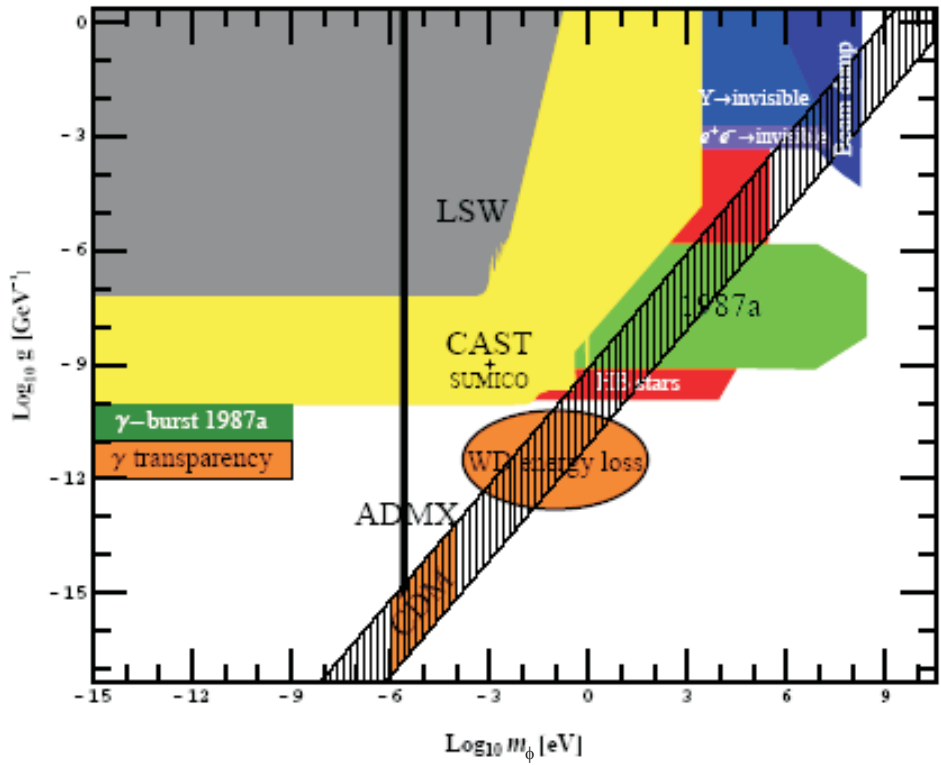}
\end{center}
\vspace{-0.6cm} \caption{\footnotesize{Summary of cosmological and astrophysical constraints for axion-like-particles (two photon
coupling $g$ vs. mass $m_{\phi}$ of the ALP). Areas with interesting astrophysical hints, e.g. the one for a non-standard
energy loss in white dwarfs~\cite{Isern:2008nt} or the one for an anomalous $\gamma$-ray transparency of the
universe (e.g.~\cite{DeAngelis:2007dy,DeAngelis:2008sk}), are marked in orange. The parameter range for the axion is shown
hatched. Note that the limit from the microwave cavity axion dark matter search experiment ADMX~\cite{Duffy:2006aa} is
valid only under the assumption that the local density of ALPs at earth is given by the dark matter density. (Compilation
from Ref.~\cite{Jaeckel:2010ni}, where also details can be found.) }} \label{fig:constraints}
\end{figure}

The sensitivity of these experiments has grown considerably over the last few years, to the point that by now they are
the most sensitive purely laboratory probes\footnote{Indeed laboratory bounds are less model dependent. They also apply
if the couplings to photons effectively depend on environmental conditions such as the temperature and matter
density~\cite{Masso:2005ym,Jaeckel:2006id,Masso:2006gc,Redondo:2008tq,Mohapatra:2006pv,Brax:2007ak,Jain:2005nh,Jain:2006ki,Kim:2007wj}.} for ALPs in the sub-eV mass range. This is shown in
Fig.~\ref{fig:constraints} which displays the present status of constraints on the two-photon coupling $g$ of ALPs versus
their mass $m_\phi$. LSW experiments currently reach a sensitivity $g\sim \mbox{few} \times 10^{-7}\,{\rm GeV}^{-1}$.
Clearly, the goal (see, e.g.,~\cite{Ringwald:2010yr,Baker:2010ma}) of the next generation of LSW experiments should be to reach and maybe even supersede the present
limits arising from astrophysics and from the non-observation of photon regeneration of solar ALPs in the magnetic field
of the CERN Axion Solar Telescope CAST (cf. Fig.~\ref{fig:constraints})~\cite{Andriamonje:2009dx}. This requires a
significant improvement in sensitivity by four orders of magnitude to the $\sim 10^{-11}\,{\rm GeV}^{-1}$ level. Beyond
entering untested parameter space, ALPs with such a coupling are particularly well motivated, both from theory as well as
phenomenology. Firstly, massless ALPs, with coupling to photons in the $g\sim \alpha/M_s\sim 10^{-11}$~GeV$^{-1}$ range
could occur naturally in string compactifications with an intermediate string scale $M_s\sim 10^9$~GeV. Secondly, there
are a number of puzzling astrophysical observations which
 may be explained by the existence of ALPs with $g$ in the above
 range~\cite{Csaki:2001yk,Csaki:2003ef,Mirizzi:2007hr,Hooper:2007bq,Hochmuth:2007hk,Payez:2008pm,Fairbairn:2009zi,Mirizzi:2009nq,Isern:2008nt,DeAngelis:2007dy,DeAngelis:2008sk,Mirizzi:2009aj,Bassan:2010ya} (the interesting areas are marked orange in Fig.~\ref{fig:constraints}.)

The most
straightforward way to increase the sensitivity for ALPs is to enlarge, $B L$, where $B$ stands for the magnitude of the
magnetic field and $L$ its length. This directly increases the probability of $\gamma\to {\rm ALP}\to \gamma$ conversion,
which at small masses scales as $\propto (BL)^4$. Nearly all the current LSW experiments (ALPS~\cite{Ehret:2009sq,Ehret:2010mh}, BFRT~\cite{Robilliard:2007bq,Fouche:2008jk}, BMV~\cite{Ruoso:1992nx,Cameron:1993mr}, GammeV~\cite{Chou:2007zzc}, LIPSS~\cite{Afanasev:2008fv,Afanasev:2008jt} and OSQAR~\cite{Pugnat:2007nu}) performed so far, recycle one or two of the long superconducting dipole
magnets from accelerator rings, such as the ones from HERA, Tevatron or LHC. A straightforward improvement can be
achieved by using a larger number of these magnets (e.g. there are about 400 superconducting dipole magnets in the
decommissioned HERA ring \cite{Ringwald:2003nsa}).

Another important step in advancing LSW experiments is the
introduction of matched optical resonators in both, production and regeneration
regions~\cite{Hoogeveen:1990vq,Sikivie:2007qm}, thereby increasing the probability for a photon to arrive at the
detector by a factor $\sim N_{\rm pass\,, 1}N_{\rm pass\,, 2}$, corresponding to the number of passes $N_{{\rm pass},i}$ in each cavity.

Implementing these advances, sensitivities in the $g\sim 10^{-11}$~GeV$^{-1}$ range, for light ALPs, seem achievable, thus opening great opportunities for discoveries.
However, both the increase in the length of the interaction region as well as the resonant cavities come at a price.
An increased length of the interaction region actually reduces the ALPs sensitivity at larger masses, and high quality cavities must have
a minimal diameter in order to avoid excessive {diffractive} losses, in turn requiring magnets with sufficiently large aperture.
It is the purpose of the present paper to discuss these effects and ways how to ameliorate them.
Moreover, we want to find an optimal configuration, based on existing technology, that maximizes the sensitivity of the experiment for a wide range
of ALPs masses.

The paper is set up as follows. In the following section we briefly recapitulate the necessary formulae for LSW experiments. In particular we review the basics of photon -- WISP -- photon oscillations and the effects of resonant cavities. In
Section~\ref{configuration} we discuss how the loss of sensitivity at large ALP masses can be ameliorated by alternating
the direction of the magnetic field (see also Refs.~\cite{VanBibber:1987rq,Afanasev:2006cv})
and leaving suitable gaps in between the magnets -- 
a previously unnoticed, but crucial feature which can be exploited, too. Section~\ref{diffraction} is devoted to the
diffractive losses in the optical cavity and the resulting requirements for the aperture of the magnets (see also~\cite{muellertalk}). We then study what is
possible with currently available magnets and what would be ideal features of future magnets.
Finally,
in Section~\ref{conclusions} we summarize our findings and conclude.

\section{Basics of photon-ALPs-photon oscillations}\label{review}

In this section we would like to recapitulate the theoretical basis of LSW experiments, in particular
to recall the general form of the oscillation probability and then deduce its explicit expression for the setup of current experiments.

At low energies, the coupling of an ALP, $\phi$, to photons can be described by the effective Lagrangian
\bb
\mathcal L=-\frac{1}4 F^{\mu\nu} F_{\mu\nu}+\frac{1}2\left(\partial_\mu \phi\partial^\mu \phi-m_\phi ^2 \phi^2
\right)-\frac{1}4{g}\, \phi F_{\mu\nu}\tilde F^{\mu\nu},
\ee
where $F_{\mu\nu}$ is the electromagnetic tensor and $\tilde F_{\mu\nu}= \frac{1}{2} \epsilon_{\mu\nu\rho\lambda}F^{\rho\lambda}$
its dual. This leads to the following equations of motion~\cite{Raffelt:1987im},
\eqb
\left[\left(\omega^2+\partial_z^2\right) {\mbox{$1 \hspace{-1.0mm} {\bf l}$}}
\ -\begin{pmatrix} -2\omega^2(n-1) &
-gB\omega
\\ -gB\omega & m_{\phi}^2 \end{pmatrix}\right] \begin{pmatrix} A \\ \phi \end{pmatrix}=0,
\label{allor}
\eqf
where $n$ stands for the refractive index of the medium through which the photon propagates.
At first order in $gBL$,
with $L$ the linear dimension associated with the extent of the magnetic field \cite{Sikivie:1983ip,VanBibber:1987rq},
one finds two possible solutions,
\bb
\phi^\pm(\vec{r},t)=e^{-i\omega t}\int d^3 r' \frac{1}{4\pi} \frac{e^{\pm ik_\phi
|\vec{r}-\vec{r'}|}}{|\vec{r}-\vec{r}'|} g\vec{E}(\vec{r}')\cdot \vec{B}(\vec{r}'), \label{mov1}
\ee
where $k_\phi=\sqrt{\omega^2-m_\phi^2}$. Specializing to the experimentally relevant case that the photon beam is sent along the $x$-axis
and that the transverse extent of the magnetic field is much larger than that of the laser, the problem becomes
one-dimensional \cite{VanBibber:1987rq},
\bb
\phi(x,t)=e^{-i\left(\omega t-k_\phi x \right)}\frac{ig}{2k_\phi} \int dx' \vec E(x')\cdot \vec
B(x').
\ee
This can be further evaluated by inserting the appropriate plane wave form 
$\vec E (\vec x,t)=\vec e_z E_0e^{i\omega\left(nx-t\right)}$ for the electric field of the laser beam, assumed to be linearly polarized in the $z$-direction, and expressing the magnetic field as $\vec B(\vec x)=\vec e_z B\,f(x)$, where $f(x)$ accounts for the spatial variation of the field,
\bb
\phi(x,t)=\frac{ig}{2k_\phi} B \omega e^{-i\left(\omega t-k_\phi x \right)} \int dx' f(x')~e^{iqx'},
\ee
where
\bb
q=n\omega-\sqrt{\omega^2-m_\phi^2}\approx \omega(n-1)+\frac{m_\phi^2 }{2\omega}
\ee
is the momentum transfer to
the magnetic field. Therefore, the probability that a laser photon converts into an axion (and vice versa) after traveling a
distance $L$ can be written as
\eqb
P_{\gamma\rightarrow\phi}=P_{\phi\rightarrow\gamma}=\frac{1}4\frac{\omega}{k_\phi}\left(gBL\right)^2
\left|F(qL)\right|^{2},
\label{prob}
\eqf
where the function $F(qL)$ is the so called form factor, defined as
\bb
F(qL)\equiv \frac{1}{L}\int_0^L dx' f(x')~ e^{iqx'}.
\label{formfactor}
\ee

So far for the general description of laser photon $\leftrightarrow$ ALPs oscillations along an arbitrary
transverse magnetic field. Now we want to specialize to the case exploited in the current generation of LSW experiments,
exploiting
single dipole magnets with a homogeneous field along the $z$-axis, both on the generation as well as
on the regeneration side. In this case,
the explicit expression of the form factor is
\bb
\left|F_{\rm single}(qL)\right|= \left|\frac{2}{qL} \sin\left(\frac{qL}2\right)\right|. \label{normallsw}
\ee
Here, the subindex ``single" is meant to stress that this expression is valid only for a
single homogeneous magnetic region.
In this case, the maximum conversion probability,
\bb P_{\gamma\to\phi } \approx
g^2B^2L^2/4,\label{maxprob}
\ee
is achieved in vacuum ($n=1$) for small momentum transfer,  $qL/2\ll 1$, corresponding to small masses,
\bb
m_\phi \ll 8.9 \times 10^{-4}~\mbox{eV} \left[\frac{\omega}{\mbox{eV}}\right]^{1/2} \left[\frac{ \mbox{m}}{L}\right]^{1/2},
\ee
where the form factor takes its maximum value, $F_{\rm single}({qL})\approx 1$.
Correspondingly, the constraints on ALPs from LSW experiments are best at small masses,
cf. Fig.~\ref{fig:constraints}. For larger masses, the
oscillations of Eq.~(\ref{normallsw}) start to dominate, and the (envelope of the) upper limit
on the coupling starts to grow as $m_\phi ^2$,
represented in the double logarithmic plot Fig.~\ref{fig:constraints} as an approximate straight line\footnote{For axion-like particles with masses of the order of the frequency of the laser $m_\phi\sim \omega$, a resonant oscillation appears and the above formulas are not valid anymore \cite{Adler:2008gk}.}.
Given the formula (\ref{prob}) it is clear that efforts should be made in the setup of LSW experiments in order to keep
$F({qL}) \sim 1$ as much as possible, for a wide range in $m_\phi\ll\omega$.

Finally, we would like to note that LSW experiments are sensitive not only to ALPs, but also to other WISPs,
notably to hidden-sector U(1)'s and mini-charged particles (MCPs)~\cite{Okun:1982xi,Ahlers:2007rd,Ahlers:2007qf,Ahlers:2008qc}
(cf. Figs.~\ref{fig:lswb} and \ref{fig:lswc}).

The interactions of the former with the standard photon are described by the low energy effective
Lagrangian
\bb
{\cal L}  \supset -\frac{1}{4}F_{\mu\nu}F^{\mu\nu}-\frac{1}{4}X_{\mu\nu}X^{\mu\nu} -\frac{\chi}{2}X_{\mu\nu}F^{\mu\nu}+\frac{1}{2} m_{\gamma^\prime}^2X_\mu X^\mu,
\ee
where $A_\mu, X_\mu$ are the electromagnetic and hidden vector potentials, $F_{\mu\nu}=\partial_\mu A_\nu-\partial_\nu A_\mu$ and $X_{\mu\nu}=\partial_\mu X_\nu-\partial_\nu X_\mu$ their respective field strength, $\chi\ll 1$ is the kinetic
mixing parameter~\cite{Holdom:1985ag}, and $m_{\gamma^\prime}$ is the hidden photon mass. Similar to the ALPs case, kinetic mixing
gives rise to photon - hidden photon oscillations. The corresponding probability can be obtained by solving the
respective equation of motion,
\eqb
\left[\left(\omega^2+\partial_z^2\right) {\mbox{$1 \hspace{-1.0mm} {\bf l}$}}
\ -
m_{\gamma^\prime}^2 \begin{pmatrix} -2\omega^2(n-1) \chi^2 &
-\chi
\\ -\chi & 1 \end{pmatrix}\right] \begin{pmatrix} A \\ X \end{pmatrix}=0,
\label{eqmhidden}
\eqf
and reads
\bb
P_{\gamma\to\gamma^\prime }=
4\chi^4 \sin^2\left( \frac{q L_{\rm tot}}{2} \right),
\label{probhidden}
\ee
where $q$ is now given by
\bb
q=n\omega-\sqrt{\omega^2-m_{\gamma^\prime}^2}\approx \omega(n-1)+\frac{m_{\gamma^\prime}^2 }{2\omega}
\ee
and where $L_{\rm tot}$ is the total distance of travel. Clearly, for photon -
hidden photon oscillations no magnetic field is needed.

Finally, the relevant equation of motion for the case of photon - massless hidden photon (the latter
kinetically mixing with the photon) oscillations  via a loop of minicharged particles (cf. Fig.~\ref{fig:lswc}) reads
\eqb
\left[\left(\omega^2+\partial_z^2\right) {\mbox{$1 \hspace{-1.0mm} {\bf l}$}}
\ + 2\omega^2{e^{2}_{{h}}}\Delta N_{i}
 \begin{pmatrix} \chi^2 &
-\chi
\\ -\chi & 1 \end{pmatrix}\right] \begin{pmatrix} A \\ X \end{pmatrix}=0,
\label{eqmhiddenmcp}
\eqf
where $e_h$ is the unit hidden sector U(1) charge, $i=\parallel, \perp$
indicate the polarization with
respect to the magnetic field, and $\Delta N_{i}$ are the magnetic field dependent, complex refractive indices describing
the refraction and absorption due to the virtual and real production of MCPs. See Ref.~\cite{Ahlers:2007rd} for details,
where also the explicit expression for the corresponding oscillation probability can be found.

\section{Refinements of LSW experiments}\label{configuration}

For the next generation of LSW experiments several important proposals have already been made to improve the sensitivity. One of the most
effective novelties is to include matched Fabry-Perot cavities in the production and regeneration sides of the
experiment~\cite{Hoogeveen:1990vq,Sikivie:2007qm}. When both cavities are tuned to the same frequency, $\omega$, it is possible to gain an enhancement
in the sensitivity for the ALP-photon coupling or the kinetic mixing parameter by the fourth root of each of the
cavities' power buildups\footnote{From now on, we chose to refer to the power buildup of the cavity, instead of the commonly used finesse {since it is the former which plays the most direct role in the production and regeneration of WISPs. See also Sec.~\ref{diffraction}.}}, $g_{\rm sens}$ or $\chi_{\rm sens} \propto (\pb_g \pb_r)^{-1/4}$.
Considering that with the available technology cavities with $\pb \sim 10^4-10^5 $ seem realistic, an improvement of the order of $10^2$ in these couplings are feasible.
The expected number of photons after the regeneration cavity of such an LSW experiment will be given by~{\cite{Hoogeveen:1990vq,Mueller:2009wt}}
\bb
N_s=\eta^2 \, \pb_g \pb_r \, \frac{\mathcal{P}_{\rm prim}}{\omega}\, 
P_{\gamma\rightarrow {\rm WISP}}^2 \tau,
\label{regphoton}
\ee
where (in a symmetric setup) $P_{\gamma\rightarrow {\rm WISP}}=P_{{\rm WISP}\rightarrow \gamma }$ is the probability of photon-ALP conversion, e.g. Eq.~(\ref{prob}) or (\ref{probhidden}), $\mathcal{P}_{\rm prim}$ is the primary laser power,  $\pb_{g,r}$ are the power build-ups of the generation and regeneration cavities,
$\eta$ is the spatial overlap integral between the WISP mode and the electric field mode~\cite{Hoogeveen:1990vq} and $\tau$ is the measurement time.

\begin{table}[t]
\centering
\begin{tabular}
{|l||l|}
    \hline
$6+6$ HERA magnets ($\ell=8.8$~m)   & $L= 52.8$~m \\
\hline
Magnetic field & $B= 5.5$~T \\
\hline
Primary laser power & ${\mathcal P}_{\rm prim}=3$~W \\
\hline
Power build-up  & $\pb_g=\pb_r=10^{5}$\\
\hline
Laser frequency & $\omega=1.17$~eV\\
\hline
Overlap between WISP mode and electric field mode  & $\eta = 0.95$ \\
\hline
Detection time  & $\tau = 100$~h \\
\hline
Dark count rate  & $n_b = 10^{-4}$~Hz \\
\hline
\end{tabular}
\caption{Benchmark values for a next generation LSW experiment.} \label{table:table1}
\end{table}

The significance of the WISP discovery scales as $S=2(\sqrt{N_s+N_b}-\sqrt{N_b})$ where $N_b$ is the expected number of background events~\cite{Bityukov:2000tt,Bityukov:1998ju}. 
To estimate the sensitivity of a LSW experiment let us assume that no significant signal over background
is found. Then a 95\% C.L. exclusion limit can be set requiring {$S<2$} which translates into $N_s<1+2\sqrt{N_b}$. 
Writing $N_s=n_s \tau$ and $N_b=n_b \tau$ with $n_s$ and $n_b$ the detection  rates of signal and background photons
we find that the limit on $n_s$ scales improves with measurement times as {$1/\tau$} if backgrounds are negligible and only as
{$2\sqrt{n_b/\tau}$} where the experiment is background dominated.
Let us consider that we are in the second situation\footnote{If feasible, it makes sense to extend the measurement period until the $1/\sqrt{\tau}$ regime is reached.}.
The corresponding expected sensitivity for the ALP-photon coupling
of such an experiment is thus
 \eqb
 \nonumber
 g_{\rm sens} &=&\frac{ {2.71} \times 10^{-11}}{\mbox{GeV}} \frac{1}{\left|F(qL)\right|} 
 \left[\frac{290 \ \mbox{Tm}}{BL}\right]
 \left[\frac{0.95}\eta\right]^{1/2}							\\
&& \times 
 \left[\frac{10^{10}}{\pb_g \pb_r}\right]^{1/4}
\left[ \frac{3 \ {\rm W}} {{\mathcal P}_{\rm prim}}\right]^{1/4}	
\left[\frac{n_b}{10^{-4} {\rm Hz}}\right]^{1/8}
\left[\frac{100 {\rm h}}{\tau}\right]^{1/8},
 \label{gmin}
 \eqf
where we have used the benchmark values for the most important parameters, 
as summarized in Table~\ref{table:table1}.

Since an improvement in the length of the magnetic region seems mandatory for future LSW experiments, we
have assumed a benchmark scenario with 6 superconducting dipole magnets on the production and the
regeneration side of the experiment (cf. Tab.~\ref{table:table1}). Correspondingly, the longitudinal
profile $f(x)$ of the magnetic field will not be so simple as in the single magnet
setups of the first generation of LSW experiments. Therefore, the form factor $F(qL)$ in
Eq.~(\ref{gmin}) has to be determined via Eq.~(\ref{formfactor}) using the actual longitudinal
profile. It is the purpose of the remainder of this section to do this for realistic arrangements of
magnets and to determine its actual influence on the sensitivity\footnote{{Clearly, the sensitivity
to the kinetic mixing between the photon and a massive hidden photon will not
depend on the longitudinal profile of the magnetic field. Still, as we will see, the laser improvements and the
increased length of the generation and regeneration sides will lead, in the next generation of LSW
experiments, to a sizable improvement in sensitivity.}}.

\subsection{Generalization to {a series of magnets with gaps in-between}}

The setups foreseen for the next generation of LSW experiments will exploit series of $N$
dipole magnets, including a natural and probably unavoidable ``gap'' with no magnetic field between each {magnet.
Therefore, let us investigate the form factor~(\ref{formfactor}) for a longitudinal profile corresponding to
$N$ equally spaced magnets, each of length $\ell$, separated from each other by a fixed length
$\Delta$. In fact, a short calculation results in
\eqb 
\left|F_{N,\Delta}(qL)\right|
&=& \left|\left(\frac{1}L \int_0^\ell dx' \ e^{iqx'}\right) \sum_{n=0}^N e^{i q n (\ell+\Delta)}\right|=
\left|\left(\frac{e^{i q\ell}-1}{qL} \right)\frac{1-e^{iqN(\ell+\Delta)}}{1-e^{iq(\ell+\Delta)}}\right| \nonumber\\
&=& \left|\ \frac{2}{qL}\sin\left(\frac{qL}{2N}\right)\frac{\sin\left(\frac{qN}{2}\left(\frac{L}N+\Delta\right)\right)}
{\sin\left(\frac{q}{2}\left(\frac{L}N+\Delta\right)\right)}\right|,
\label{fgaps}
\eqf
with $L=N\ell$ the total length of the magnetic field. 

The interpretation is clear. The form factor is the sum of the individual form factors of the production in the different 
magnets (which being equal factorizes out), weighted by a phase that accounts for the phase delay of the ALPs produced 
in the different magnets (Since a common phase is arbitrary, we have measured these phases with respect to the axions produced in the first magnet, so it is $e^{i q n(\ell+\Delta)}$ for the ($n+1$)th magnet).

This function has two types of zeros. First, it is evident the zeros of the form factor of one single magnet ($q\ell =2\pi k, k\in \mathbb Z^+$) should be still there since it appears as a common factor. Secondly, there are the zeros related to the gaps, appearing when $qN(\ell+\Delta)=2\pi k', k'\in \mathbb Z^+$ \emph{and} $q(\ell+\Delta)\neq 2\pi k'', k''\in \mathbb Z^+$. For non-zero $\Delta$ these conditions are fulfilled when
$k'$ is \emph{not} an integer multiple of $N$, i.e. $k'=1,2,...,N-1,N+1,...2N-1,2N+1$, etc. This means that for each zero of the single magnet form factor there are $N-1$ zeros coming from the gaps.
When $\Delta\to 0$ the combination of the two types of zeros coincide with the zeros of a magnet of length $L$.
As a check, we note that Eq.~(\ref{fgaps}) approaches the previous single magnet result~(\ref{normallsw}) in the limit
$\Delta\rightarrow 0$ or $N=1$, as it should.

All in all, the presence of gaps does not alter the number of zeros of the form factor, which ultimately means dips in sensitivity we want to avoid. It does however modify the values of $q$ and hence the ALP mass where they appear. 
As we increase $\Delta$ from 0 the zeros displace towards smaller values of $q$ (and hence of $m_\phi$).
In particular, the first zero of the gapless form factor in Eq.~(\ref{normallsw}), $q=2\pi/L$, now moves to $q=2\pi/(L+N\Delta)$.
This means that ultimately, the price to pay for enlarging the magnetic region by piling up individual magnets with gaps in between is the loss of the coherence of the photon-ALP conversion already at smaller masses (with respect to the ideal gapless case).

Not surprisingly, our result coincides with that obtained by introducing phase shift plates inside the magnetic region~\cite{Jaeckel:2007gk}. 
In~\cite{Jaeckel:2007gk}, the authors argue that it is possible to put the photon and ALP waves in phase after some distance by the insertion of a phase shift plate that advances or delays the photon phase with respect to the axion wave. In our case, we can do the same after 
each magnet by choosing the appropriate gap length\footnote{We note, however, that this is only true for ALPs. For hidden photons, which do not interact with the magnetic field, it is only the total length (including the gaps) that counts.}.
The clear advantage is that the gaps do not introduce extra optical losses. 


\begin{figure}[t]
\begin{minipage}{0.6 \textwidth}
\centering
\includegraphics[width=1\textwidth]{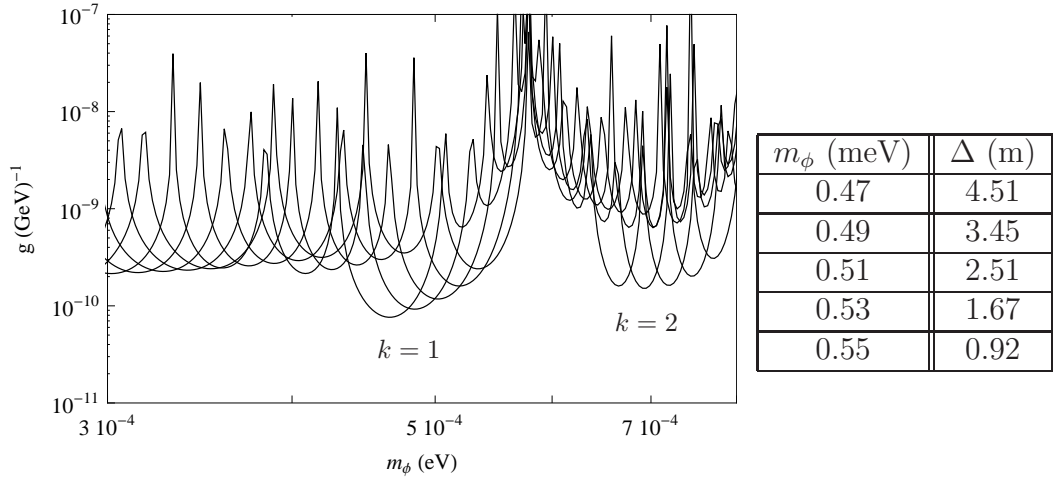}
\Text(-50,60)[l]{ \footnotesize{$k=2$}}
\Text(-140,50)[l]{ \footnotesize{$k=1$}}
\label{fig:gapscurves}
\end{minipage}
 \   \ \hfill \begin{minipage}{0.4 \textwidth}
\begin{tabular}{|c||c|} \hline
${m_\phi}$ (meV)  &  $\Delta$ (m)  \\
\hline
{0.47} & {4.51} \\
\hline
{0.49} & {3.45} \\
\hline
{0.51} & {2.51}\\
\hline
{0.53} & {1.67}\\
\hline
{0.55} & {0.92}\\
\hline
\end{tabular}
\end{minipage}
\vspace{-0.5cm}
\caption{\footnotesize{Sensitivity
for an LSW experiment using $6+6$ HERA magnets. The considered {benchmark parameters of the
setup are} given in Tab.~\ref{table:table1}. The
enhanced zone
corresponds to $m_\phi=\ 4.7-5.5 \ \times 10^{-4}$~eV,
and the size of the respectively needed gaps are shown in the table.
The $k$'s are the corresponding integers in Eq.~(\ref{maxi}).
}}
\end{figure}

Since the gaps' length appears as a free parameter which is adjustable to our needs (besides the number of
magnets, which can not be altered significantly), we can optimize the form factor Eq.~(\ref{fgaps}) for a given non-zero ALP mass 
by choosing the appropriate value of $\Delta$. The maximum of Eq.~(\ref{fgaps}) happens when the ALP production in all the magnets
adds up in phase, that is when the phases in the sum in  Eq.~(\ref{fgaps}) add up coherently, i.e. when
\bb
\frac{q\ell}{2}\left(1+\frac{\Delta}\ell\right)=k\pi, \quad \quad (k \in \mathbb{Z^+}) .
\label{maxi}
\ee
At these maxima, the form factor is reduced to the one corresponding to a single magnet of
length $\ell$, ${F_{N,\Delta\mid_{\rm max}}}=F(q\ell)$. This is the best one can achieve
with a multimagnet configuration without using phase-shift plates or filling the oscillation region with a buffer gas~\cite{Ehret:2010mh}.

However, the fact that this optimization is mathematically possible does not mean that 
it can be conveniently implemented in a realistic experiment.
As we can easily see, if we want to optimize for instance the low mass region 
$q \ell/2<1$ we shall have at least $q\ell (1+\Delta/\ell)/2=\pi$, requiring $\Delta>\ell$, which is a very large gap.
In Sec.~\ref{diffraction} we will study the limitations on the length $L$ imposed by the current available magnets and will
understand that the total length is a valuable good\footnote{Indeed in some cases it might even be useful/possible to fill the unavoidable 
gap with another magnet.}.

We conclude that changing the gap size allows for filling the sensitivity dips of the form factor, improving the sensitivity of LSW 
experiments especially at large masses corresponding to $q\ell/2> 1$. For the {benchmark setup proposed in Tab.~\ref{table:table1}, this allows to significantly improve the sensitivity in
the mass range} $m_\phi \sim 4 \times 10^{-4}-10^{-3}$~eV.
{If we try to extend this down to smaller masses, the size of the gap becomes excessively large for a realistic experimental setup.
For instance, in order to maximize the sensitivity for an ALP mass of $m_\phi=10^{-4}$~eV, an enormous gap between the
HERA type ($\ell=8.8$~m) magnets of $\Delta=285$~m would be needed.
As an illustration, we show in Fig.~\ref{fig:gapscurves} the projected sensitivity for a $6+6$ HERA magnet configuration,
taking into account gaps between the magnets. The enhancement corresponds to an ALP mass range
of (4.7 - 5.5)$\times 10^{-4}$~eV. The input details of the configuration are the same as the ones shown in Eq.~(\ref{gmin}).
We have chosen values of $\Delta$ that are reasonable experimentally. }

\subsection{Generalization to {a series of magnets with alternating} polarity}

Now we will extend this study considering an array on $N$ identical magnets of length $\ell$, segmented into $n$ subgroups of alternating polarity, such that the total magnetic length is given by $L=N\ell$.
This idea was proposed in {Ref.~\cite{VanBibber:1987rq}} as a technique to peak the form factor at nonzero values of $q$. The formula, neglecting the gap between the magnets is given by~\cite{VanBibber:1987rq,Afanasev:2006cv}
\eqb
\left|F_n (qL)\right| = \left\{
\begin{array}{l l}
\left|\frac{2}{qL}\sin\left(\frac{qL}{2}\right)\tan\left(\frac{qL}{2n}\right)\right|{,} & \quad \text{$n$ even}{,}\\
\left|\frac{2}{qL}\cos\left(\frac{qL}{2}\right)\tan\left(\frac{qL}{2n}\right)\right|{,} & \quad \text{$n$ odd}{.}\\
\end{array} \right.
\label{alterold}
\eqf
The maxima of this form factor (with respect to $qL/2$) happen close to the poles of the tangent, which are at the same time
zeros of the accompanying {sine or cosine}. 
The presence of the external factor $2/(qL)$ slightly moves the maxima away from the poles of the tangent, this 
correction disappearing at large $n$ (corresponding to large $qL/2$). The absolute {maximum} is the one
with smallest $qL/2$.
We have searched numerically the absolute maxima of the form factor in Tab.~\ref{tab:maxima} for different values of $n$.
For sufficiently large $n$ we can treat the the problem analytically performing an expansion in $1/(qL)$. 
One can see that the solutions approach asymptotically $qL/2=n\pi/2$, and for them ${F_n(qL)}\rightarrow 2/\pi$.

\begin{table}[t]
\caption{First maxima of the form factor for an array of magnets alternating in $n$ subgroups.
For $n\to \infty$, $qL/2\to n \pi/2$.}
\begin{center}
\begin{tabular}{|c|c|c|c|c|c|c|}
\hline
$n= $ & $1$ & $2$ & $3$ & $4$ & $5$ & $6$ \\			\hline
$qL/2=$ & $0$ & $2.331$ & $4.131$ & $5.832$ & $7.486$ & $9.115$\\ 			\hline
\end{tabular}
\end{center}
\label{tab:maxima}
\end{table}%

If we now include a fixed gap in-between the magnets, as we did
in Eq.~(\ref{fgaps}), we find 
\eqb
\left|F_{N,n,\Delta}(qL)\right|=\left\{ \begin{array} {l l}
\left|\frac{2}{qL} \sin\left(\frac{qL}{2N}\right) \frac{\sin\left(\frac{qN}2(L/N+\Delta)\right)}{\sin\left(\frac{q}{2}(L/N+\Delta)\right)}
\tan\left(\frac{qN}{2n}(\frac{L}N+\Delta)\right)\right|, \ \ \ \mbox{$n$ even}{,}
\label{alternating1}\\
\left|\frac{2}{qL} \sin\left(\frac{qL}{2N}\right) \frac{\cos\left(\frac{qN}2(L/N+\Delta)\right)}{\sin\left(\frac{q}{2}(L/N+\Delta)\right)}
\tan\left(\frac{qN}{2n}(\frac{L}N+\Delta)\right)\right|, \ \ \ \mbox{$n$ odd}{.}\\
\end{array}\right.
\label{alternating2}
\eqf
In the limit $\Delta \rightarrow 0$, formulas~(\ref{alterold}) are recovered. 
Moreover, the non-alternating form factor (Eq.~\eqref{fgaps}) is recovered for $n=1$.

Depending on the choice of the pair
{$\left\{N,n\right\}$,} there is a different maximum for the form factor.
{Indeed, they can be searched as before by finding the poles of the tangents. In general we find that 
they are close to $N q(\ell+\Delta)/2=n(2k+1)\pi/2, k\in \mathbb Z^+$ and the most important one corresponds to 
$k=0$. This is asymptotically true as $N q(\ell+\Delta)/2$ becomes large but again, the poles have to be searched
for numerically for ${\cal O}(1)$ values.} 
In general and as with the previous case, the maxima displace towards smaller $q$ when we increase $\Delta$.

We can again use the gap length $\Delta$ to optimize the form factor for a given ALP mass. 
The absolute maxima of Eq.~\eqref{alternating2} cannot be found analytically except in particular cases, 
however it is easy to see that they lie either very close to the poles of the tangent or to the zeros of the sine 
in the denominator (only for the $n$-odd case). 
The poles of the tangent are   
\bb
\label{altmaxi}
\frac{q N}{2 n}(\ell+\Delta) = \frac{\pi}{2}+\pi k,  \ \ \ \ \ k \in
{\mathbb{Z^+}}, 
\ee
and for these values the form factor evaluates to 
\bb
F_{N,n,\Delta}(qL) = F_{\rm single}(q\ell)\frac{n/N}{\sin\frac{n}{N}\left(\frac{\pi}{2}+\pi k\right)} .
\ee
The zeros of the sine in the denominator are given by Eq.~\eqref{maxi} (they appear also in the non-alternating configuration)
and for them the form factor evaluates to $F_{N,n-odd,\Delta}(qL) = F_{\rm single}(q\ell)/n$.

The $N=n$ case (i.e., a $ \uparrow \downarrow
\uparrow\downarrow...$ configuration) can be solved analytically. 
One finds that the absolute maxima coincide \emph{exactly} with the poles of the tangent.
In this particular case, full coherence between the ALP production in all the magnets can be achieved and correspondingly
the form factor saturates to the single magnet form factor $F_{N,N,\Delta}(qL) = F_{\rm single}(q\ell)$
(In the second set of maxima, $F_{N,N,\Delta}$ is suppressed by a factor $1/N$).

The case $n=1$ is trivial since it reduces to the non-alternating configuration. In this case, 
as we have said one can obtain also $F_{N,1,\Delta}(qL) = F_{\rm single}(q\ell)$ but this happens 
for the zeros of the sine in the denominator and the poles of the tangent produce non-absolute maxima
for which the form factor is a bit smaller (is easy to see that $F_{N,1,\Delta}(qL) \to F_{\rm single}(q\ell)\times 2/\pi$ in the $N\to \infty$ limit). 

All in all we see that for certain ALP masses, the sensitivity can be nearly fully restored.
Also, it is easy to see that this optimization can be performed even in the case $q\ell/2\leq1$, 
unlike the non-alternating configuration considered in the previous section. 
As an example, 
in the case of $N=n$ HERA magnets, with gaps of $\Delta =1$~m, the first maximum condition, $k=0$,
is reached for $m_\phi=3.87\times 10^{-4}$~eV, a region
where the previous setup (without alternation and gaps) had no sensitivity, cf. Fig.~\ref{fig:gapscurves}. 


\subsection{Some examples}
As an example, let us consider a possible $6+6$ HERA configuration, {i.e.} 6 magnets in the production and in regeneration side, respectively,
with the {benchmark} parameters given in Tab.~\ref{table:table1}.

With six magnets on each side of the experiment we can produce four different symmetric configurations: $\uparrow
\uparrow\uparrow\uparrow \uparrow\uparrow$, $\uparrow \downarrow\uparrow \downarrow\uparrow \downarrow$,
$\uparrow\uparrow \downarrow\downarrow\uparrow\uparrow$, $\uparrow\uparrow\uparrow\downarrow\downarrow\downarrow$.
In {Fig.~\ref{fig:examples} a)} we display the sensitivity of all four possible $6+6$ configurations, taking
into account seven different gap sizes. We infer that, using these alternating field configurations, we are able to restore the sensitivity almost completely
up to masses of order $m_\phi\sim 5\times 10^{-4}$~eV.

\begin{figure}
\centering \subfigure[]{\label{fig:fplot}\includegraphics[width=0.45\textwidth]{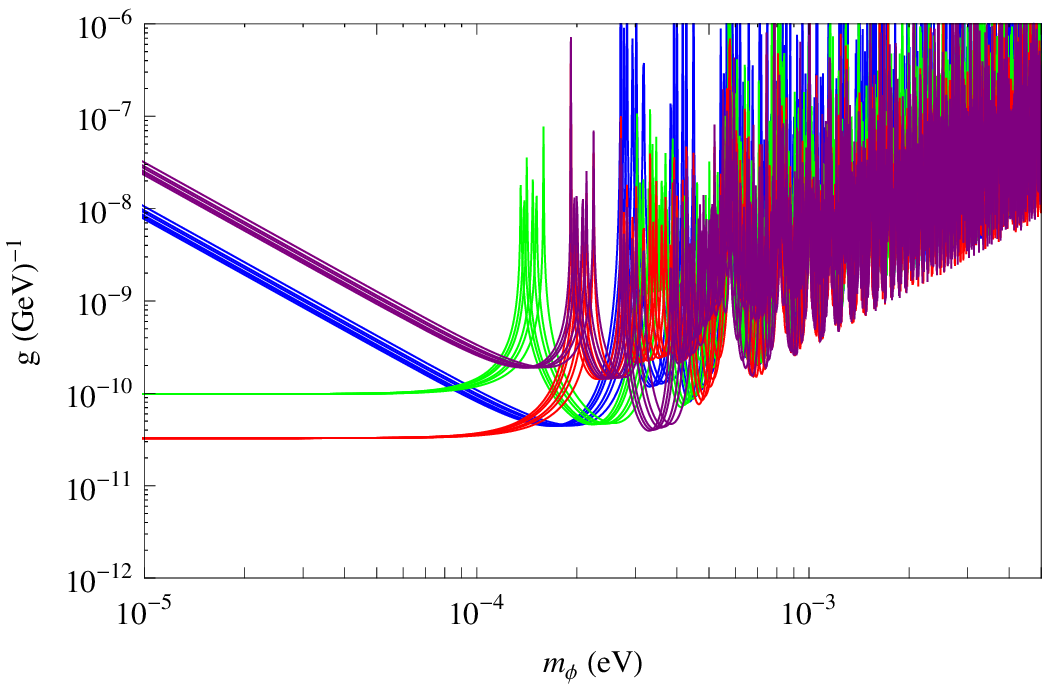}}
\Text(-160,30)[l]{ \tiny{$6 \times \left( 5.5 \mbox{~T}, \, 8.8 \mbox{~m}\right)\times 2$}}
\Text(-140,120)[l]{ \tiny{\color{red}{$\uparrow\uparrow\uparrow\uparrow\uparrow\uparrow$}}}
\Text(-140,110)[l]{ \tiny{\color{blue}{$\uparrow\uparrow\uparrow\downarrow\downarrow\downarrow$}}}
\Text(-140,100)[l]{ \tiny{\color{green}{$\uparrow\uparrow\downarrow\downarrow\uparrow\uparrow$}}}
\Text(-140,90)[l]{ \tiny{\color{purple}{$\uparrow\downarrow\uparrow\downarrow\uparrow\downarrow$}}}
\subfigure[]{\label{fig:splot}\includegraphics[width=0.45\textwidth]{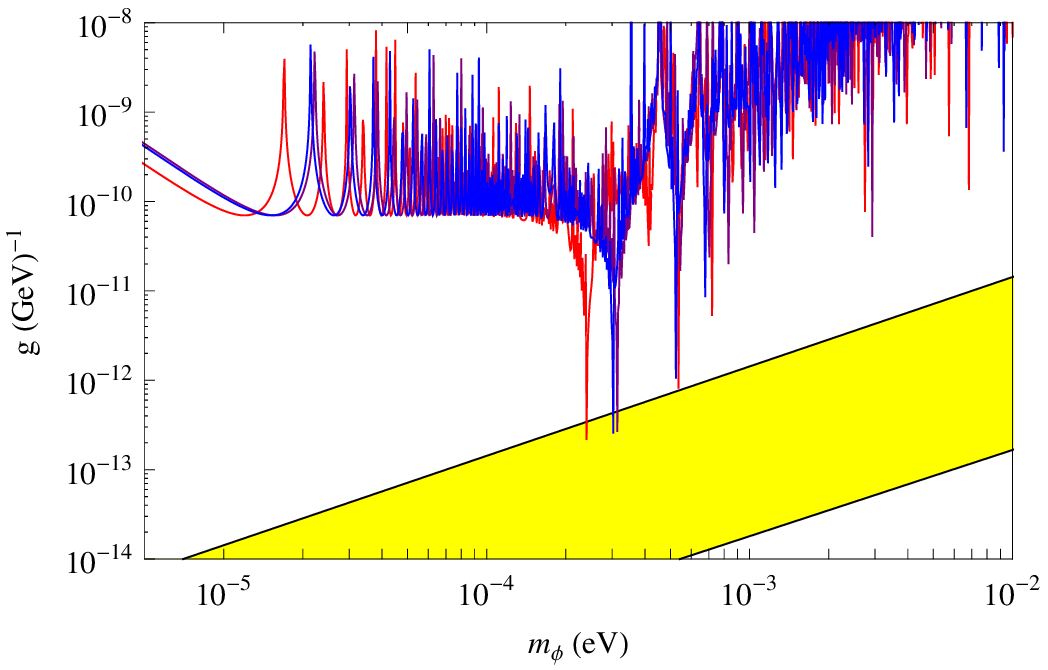}}
 \Text(-175,80)[l]{ \tiny{\color{purple}{$\Delta=0.6$~m}}}
 \Text(-175,70)[l]{ \tiny{\color{blue}{$\Delta=0.9$~m}}}
 \Text(-175,60)[l]{ \tiny{\color{red}{$\Delta=11.2$~m}}}
\Text(-175,30)[l]{ \tiny{$400\times \left( 9.5 \mbox{~T}, \, 14.3 \mbox{~m}\right)\times 2$}}
\Text(-75,52)[l]{ \tiny{\bf{Axion models}}}\caption{
\footnotesize{
a)``$6+6$'' configuration of HERA magnets. Combining all the possible wiggler configurations it is
possible to restore the sensitivity of the experiment until the first zero of the form factor,
where $qL/2N=\pi/2$. The size of the gaps are $\Delta [m]=\left\{4.5,3.9,3.4,2.5,1.9,0.9\right\}$. 
b) Sensitivity of a toy configuration of ``$400+400$'' LHC magnets in order to touch the predictions for the QCD axion (``axion band").
Using a gap in-between the magnets of $\Delta=0.6$~m we are able to scratch the axion band in $m_\phi\sim 3.15 \times 10^{-4}$~eV. In order to scan an ALP mass of $m_\phi  \sim 2.4 \times 10^{-4}$~eV, a gap of $\Delta=11.2$~m is needed. 
However, in order to get the sensitivity shown in this plot the bore aperture radius of the magnet should be at least $16$~cm.}} \label{fig:examples}
\end{figure}

The gaps in the sensitivity, where $qL/2N=k \pi$, with {$k \in \mathbb{Z^+}$,}
can not be covered with this setup, since all the form factors {are zero} there.
However, {the} ALPS Collaboration~\cite{Ehret:2010mh} was able to fill {such} regions
introducing a gas inside their {cavity, changing the refractive index in the generation and regeneration zone.}
Another possibility to cover {these} regions is to use phase shift plates, or a
movable wall~\cite{Chou:2007zzc}.

An important issue for the next generation of LSW experiments would be to know their potential to achieve sensitivity in
the parameter range of the proper {QCD} axion.
In Fig.~\ref{fig:examples} b) we show a (way too optimistic) toy example in order to answer this question: We have chosen an
alternating {field} configuration, with $N=n$.
According to Eq.~(\ref{altmaxi}), we select the appropriate gap to peak in the
mass range $m_\phi \approx 3 \times 10^{-4}$~eV. We used
400+400 magnets, with a magnetic field of 9.5 Tesla and a length of $\ell=14.3$~m each.
We see that even this overly optimistic configuration can just scratch the QCD axion!

Clearly, this setup is unrealistic: it needs more than 400 LHC magnets ($\ell=14.3$~m, $B=9.5$~T) in each the production and
the regeneration side of the experiment, enlarging the size of the cavity to  $\sim 5960 $ meters!
As we will see in the next section this entails, that in order to assume the benchmark values of Table~\ref{table:table1},
magnets with an aperture bore radius of $16$~cm are needed. This is considerably more than the $3$~cm of the existing HERA magnets.

In short, with the current designs LSW experiments will enter an unexplored
 region in parameter space, but are still far from being able to test the QCD {axion\footnote{{Coherence is kept up 
 to higher masses in LSW experiments exploiting more energetic photons, e.g. from synchrotron
radiation sources or X-ray free-electron lasers~\cite{Rabadan:2005dm,Dias:2009ph}. 
However, currently the sensitivity is far away from the QCD axion 
because the average photon flux of current or planned facilities is many orders of magnitude smaller than the
one reachable with optical lasers {(for a pioneering experiment in this respect see Ref.~\cite{Battesti:2010dm})}. 
Besides, an important disadvantage is the impossibility to set-up cavities.}}.}

\section{Expected sensitivity for next generation of LSW experiments}\label{diffraction}

In the preceding section we have discussed the possibility {to} change the length between the magnets as a technique
to improve the {sensitivity} of the experiment for certain ALP masses.
Now we would like to make the discussion more realistic as well as investigating the possibilities for the next
generation of photon regeneration experiments, given the actual equipment.
{For this, the maximum
length} of the cavity is crucial, since {on} the one hand the probability of photon-ALP conversion for small masses scales as
$L^4$, {and on the} other hand, bigger cavities will give us more freedom to choose {gaps} between the magnets.

{\subsection{Diffractive losses}}

\begin{figure}
\centering
\includegraphics[width=0.7\textwidth]{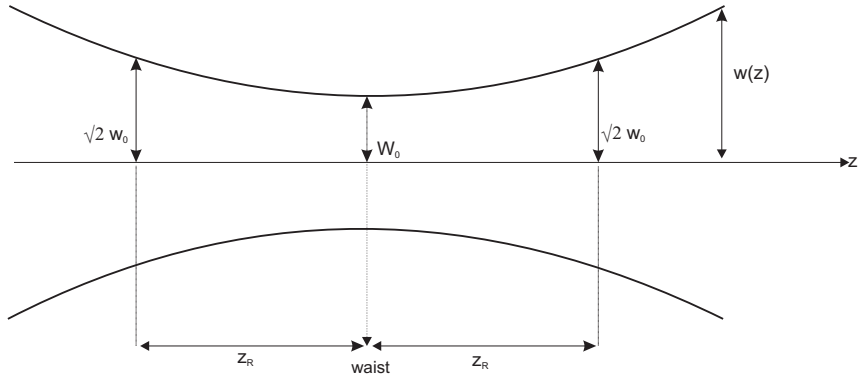}
\caption{\footnotesize{Collimated waist region of a Gaussian beam. The {Rayleigh}
range is defined as the distance which the beam travels from the waist before the beam area doubles. It is {the characteristic} distance before the beam begins to diverge significantly.}}
\label{fig:rayleigh}
\end{figure}
Concerning the cavity, we can estimate the approximate optimum length for the different superconducting dipoles available.
The bore aperture of the magnets is a key factor in order to determine the length of the cavity.\\
{Let us consider} a collimated Gaussian beam, propagating in the $z$ direction. The point at which the beam has a planar wave front with
curvature $R=\infty$, is known as the beam waist, and it is characterized by a particular spot size, $w_0$. The power intensity of a beam
falls off very rapidly with the distance, and at some arbitrary length $z$, the spot size can be described by $w(z)$,
\bb w(z)=w_0\sqrt{1+\left(\frac{z}{z_R}\right)^2}.\ee 

The distance $z_R$ is known as the Rayleigh range and defines the point at which the beam
area doubles. {It is} given by $z_R={\pi w_0^2}/{\lambda }$, where $\lambda$ is the wavelength of the laser in the medium. 
The sketch of the configuration is shown in {Fig.~\ref{fig:rayleigh}}.
The radial intensity variation of a Gaussian beam with spot size $\omega$ is given by
\bb I(r)=\frac{2{\cal P}}{\pi w^2} e^{-2r^2/w^2}, \ee
 where ${\cal P}$ is the total power of the optical beam at the position of the waist.
The fraction of the power transmitted through a circular aperture of radius $a$ is given by
\bb
\frac{\int_0^a I(r)r dr}{\int_0^\infty I(r)r dr}
= 1-e^{-2 a^2/w^2}. \label{power}
\ee

The peak amplification factor of the incident power laser 
inside an optical Fabry-Perot cavity -- known as power buildup~($\pb$) -- occurs at the resonance condition, $\omega={k\pi}/{L}$,
where $k\in \mathbb{Z}$, and $L$ is the length of the cavity, and is given approximately by~\cite{siegman}
\bb 
\pb \approx \frac{4\delta_1}{\left(\delta_0+\delta_1+\delta_2\right)^2}, 
\ee 
where $\delta_0$ accounts for the round-trip losses in the cavity and $\delta_{1,2}$ are the losses
due to the transmissivity of the cavity mirrors\footnote{The formal definition of the delta coefficients is $R_i\equiv e^{-\delta_i}$, where $R_i$ stands for the
reflectivity of the $i$-th mirror. The coefficient $\delta_0$ is related to the internal losses inside the cavity, therefore $\delta_0=4\alpha_0 L$
where $\alpha_0$ is the attenuation coefficient in the cavity.}.
In order to maximize $\pb$ we have to keep $\delta_0$ and $\delta_2$ \emph{as small as possible}. {Then} the optimum value for $\delta_1$ is $\delta_1=\delta_0+\delta_2$ corresponding to a so called {\it{impedance-matched}} cavity ($\pb=(\delta_0+\delta_2)^{-1}$). 
Let us remark that for an impedance-matched cavity, a simple relation can be established between the power build-up and the cavity's finesse
\bb \mathcal F = \pi \pb  . \ee
For a cavity in good vacuum conditions, {one of the dominant contributions} to $\delta_0$ arises from clipping the cavity mode.
If the beam waist is located near one of the mirrors, then $\delta^{\rm clip}_0=e^{-2 a^2/w^2(Z)}$ where $Z$ {is} the distance of the waist to the 
cavity end mirror. If the waist {lies} in the center the clippling occurs near both mirrors and a factor of 2 arises. 
We shall only consider the former case, which is more advantageous since it optimizes the spatial matching of the generation and regeneration cavity modes~\cite{Hoogeveen:1990vq}. 
{Other contributions to $\delta_0$ like non-Gaussian beam profiles, imperfections in the mirrors, etc., can be subsummed in an additional parameter $\delta^*_0$. The laser frequency will be also crucial to achieve maximum sensitivity:
if it is too large, dispersion in the mirrors might increase the roundtrip losses.
For this reason we have chosen to use $1064$~nm wavelength light all throughout the paper. The experience gained in the gravitational 
wave community in building high finesse cavities containing extremely intense beams with this type of light makes this choice most practical~\cite{grav}.}

Let us now describe one possible optimization procedure for the design of a LSW experiment. We start with a number of dipole magnets of field strength $B$ and aperture $a$ and ask for the parameters of the cavity that optimize our sensitivity for the ALP coupling $g$.
Let us focus in the low mass range, $m_\phi\ll 10^{-4}$ eV, larger masses will be {dealt with later on}.
Our final sensitivity to the coupling constant $g$, given by eq.~\eqref{gmin}, depends mainly on the combination
\bb
\label{opti}
 g_{\rm sens}\propto \frac{1}{L{\pb^{1/2}}}= \frac{1}{L}\left(e^{-\frac{2 \pi^2 a^2 w_0^2}{\pi^2 w_0^4+(Z\lambda)^2}}+{\delta_0^*}+\delta_2\right)^{1/2}
\ee
where we have used the impedance matching condition and substitute $z_{\rm R}$ in terms of $w_0$. 
We have taken the generation and regeneration cavities to be equal, $\pb_g=\pb_r=\pb$,  
to maximize the bandwidth overlap.

The above function can be easily minimized.
First it is clear, that the highest sensitivity is reached for} \emph{the lowest possible value for} ${\delta_0^*}+\delta_2$.
Currently the best mirrors are in the  $\delta_2\sim 10^{-5}$ ballpark so we will keep this number for the rest of the paper.
{We will also assume that $\delta_0^*$'s smaller or of the same order can be achieved.}
We have now three remaining free parameters, $L,Z$ and $w_0$. 
Optimization with respect to $w_0$ is trivial, in order to render the exponential as small as possible the argument has to be maximum and this happens at $w_0^2= Z\lambda/\pi$ (i.e. at $z_{\rm R}=Z$).

In order to minimize with respect to $Z$ and $L$ we first substitute the condition $z_{\rm R}=Z$ in Eq.~\eqref{opti},
\bb
\frac{1}{L}\left(e^{-\frac{\pi a^2}{Z \lambda}}+{\delta_0^*}+\delta_2\right)^{1/2}.
\ee
{Note that $Z> L$ since the length of the cavity $Z$ has to encompass the magnetized length $L=N\ell$
plus the length of the gaps, $N\Delta$, plus maybe a {field free} zone near the mirrors of the cavity, called $dz$.}
There is a certain difficulty in defining properly a minimization since $Z$ and $L$ are related but $Z$ can grow continuously and $L$ only in multiples of the magnet length $\ell$. {To start with let us ignore these difficulties and just consider an $L\simeq Z$ which can be varied continuously,} i.e. 
$N\Delta+dz$ is small (we want to have as {long a} magnetic region inside the cavity as we can). 
We find the minimum with respect to $Z$ to be given implicitly by\footnote{{As it turns out this equation even holds if we take $Z=N(\ell+\Delta)$ and optimize with respect to the number of magnets, $N$.}} 
{\bb
\label{osti}
e^{-\frac{\pi a^2}{Z \lambda}} \left(\frac{\pi a^2}{2 Z \lambda}-1\right)={\delta_0^*}+\delta_2 .
\ee
}

\begin{figure}[t!]
\centering
\includegraphics[width=0.45\textwidth]{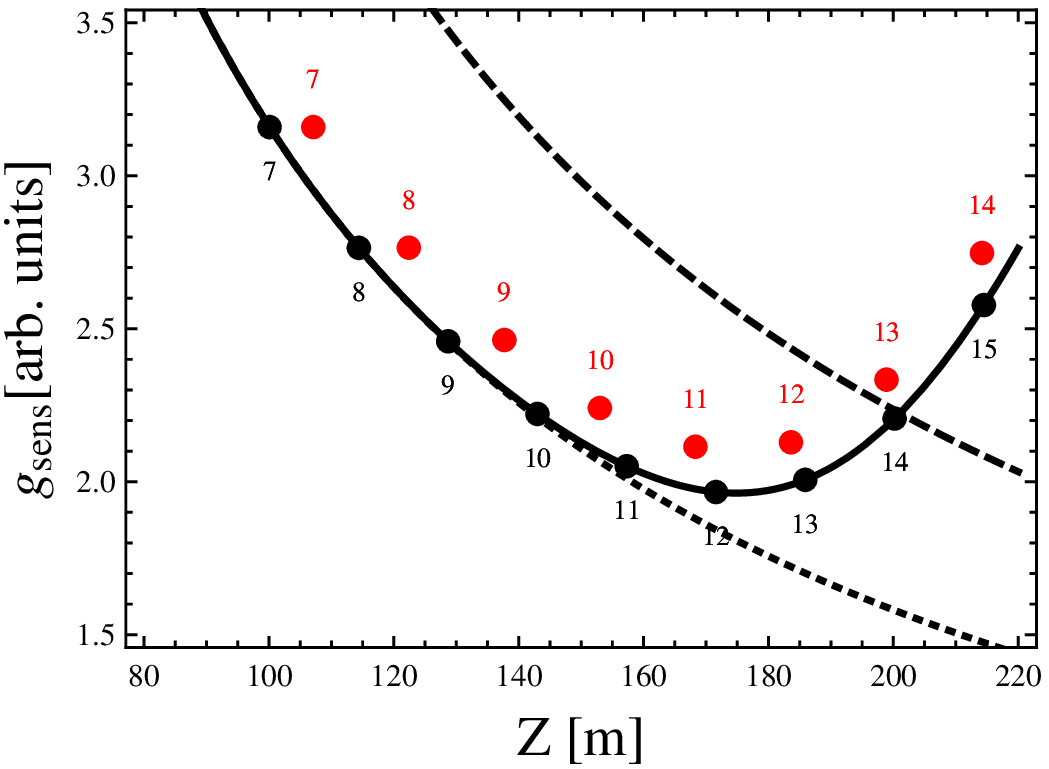}
\includegraphics[width=0.45\textwidth]{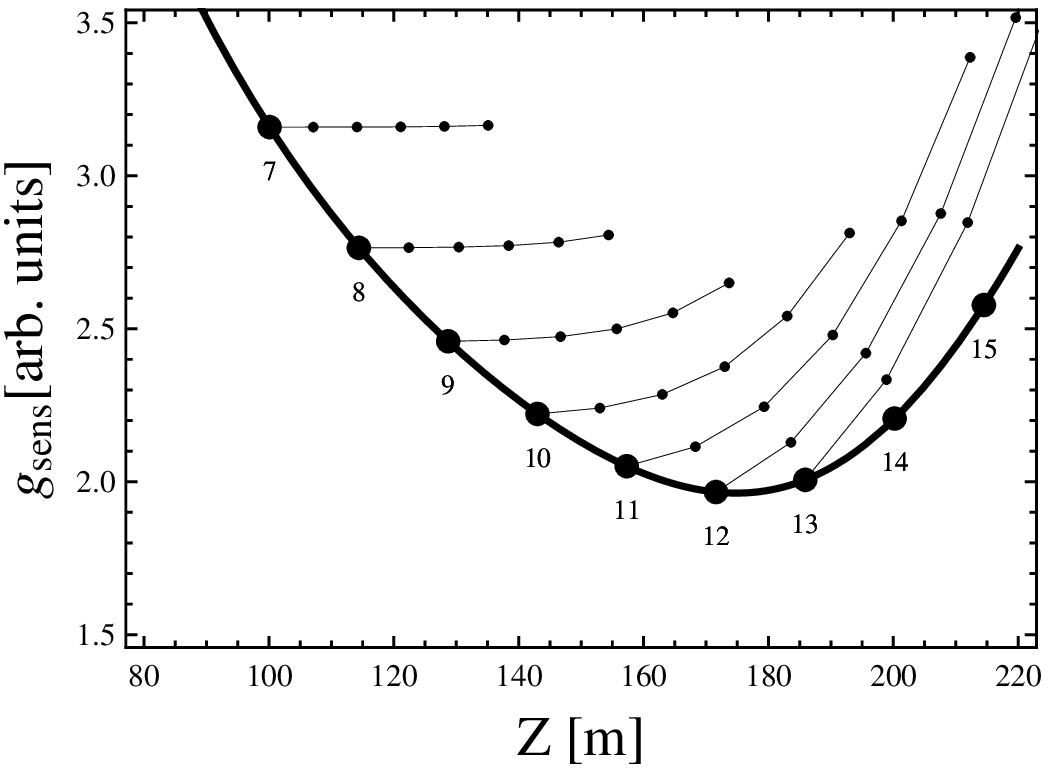}
\caption{\footnotesize{The quantity $1/(L{\pb}^{1/2})$ to minimize in the design of a LSW experiment looking for 
ALPs (in arbitrary units; {and for $m_{\phi}=0$})
as a function of the total length $Z$ of the generation or regeneration cavity. 
We have particularized for the case of the LHC with $a=28$ mm.
The solid line shows the optimal case in which the whole cavity length, $Z$, can be filled with magnets $Z=L$.
The black dots indicate the points in which $L$ is an integer multiple of the LHC magnetic length $\ell=14.3$ m, i.e. the possible realistic configurations with negligible gap size. 
LEFT: The red points show the configurations when the gap between the magnets increases to $\Delta=1$ m, ($Z=N(\ell+\Delta)$). 
The dotted line indicates an ideal case in which roundtrip losses (including clipping) would be negligible, $\delta_0=0$ or $a=\infty$. 
The dashed line assumes $\delta_0=\delta_2$.
RIGHT: The thin lines emerging from each gapless configuration to the right and up represent configurations with gap size $\Delta=1,2,3,4,5$ m.
}
\label{fig:opti}}
\end{figure}

This equation can be solved numerically but an approximate solution is very easy to find. 
Ignoring the parenthesis for a while, the equation has the solution $Z=-\pi a^2/(\lambda \log ({\delta_0^*}+\delta_2))$. 
This solution makes the parenthesis $-\log\sqrt{{\delta_0^*}+\delta_2}-1\sim 4.75$ ({taking} $Z=L$). The real solution 
will be a bit smaller such that the exponential kills this additional factor.  
As an example, for ${\delta_0^*}+\delta_2=10^{-5}$, the simple estimate gives $0.0869 \pi a^2/\lambda$ while numerically 
we find the optimum to be
\bb
\label{miniformula}
Z_{\rm opt}=0.0755 \frac{\pi a^2}{\lambda}=89.2\ {\rm m} \left(\frac{a}{20\ {\rm mm}}\right)^2 \frac{1064\ {\rm nm}}{\lambda},  
\ee 
{which corresponds to $\pi a^2/(Z_{\rm opt}\lambda)\simeq 13.25$ and then to ${\delta_0^{\rm clip}/(\delta_0+\delta_2)}=0.177$.
In previous literature~\cite{Mueller:2009wt} a value ${\delta_0^{\rm clip}/(\delta_0+\delta_2)}\sim 1$ was used, leading to a slightly less optimal setup
(see the dashed line in Fig.~\ref{fig:opti} (right)).}

Let us summarize our optimization procedure. First we have optimized with respect to $\delta_2$, the smallest value always results in an optimum sensitivity. Then, the impedance matching condition gives an optimum for $\delta_1=\delta_0+\delta_2$. 
Since $\delta_0$ {encompasses} the clipping losses it is a function of the length of the cavity and has to be maximized together with the
total magnetic length. Equation~\eqref{osti} tells us precisely the optimum cavity length $Z_{\rm opt}$ taking into account 
these relation and has to be solved numerically by specifying a value for $\delta_0^*+\delta_2$ and a relation between $L$ and $Z$ (assumed $L\sim Z$).

Now we can tackle the fact that $L$ can only grow by multiple integers of $\ell +\Delta$. 
This can be numerically computed. In Fig.~\ref{fig:opti} (left) we have plotted as a black solid line 
the quantity to minimize (Eq.~\eqref{opti}) as function of $Z$ using $L=Z$ for an array of LHC magnets for which $a=28$ mm. 
The black dots represent the value of Eq.~\eqref{opti} as a function of the 
number of magnets, i.e. $Z=L=N\ell$. The minimum of the continuous line stands clearly around $Z\sim 175$ m (our minimization formula, Eq.~\eqref{miniformula}, gave $Z_{\rm opt}=174.78$ m). The possible realization closer to it corresponds to 
an array of 12 LHC dipoles. Note however that the minimum is reasonably wide and there is some freedom to chose $N$
without spoiling the sensitivity too much.

{Let us now move on to discuss the high mass range. 
The setup described above maximizes the sensitivity at $m_\phi\to0$ (equivalently $q\to 0$) but will have some unavoidable dips in sensitivity corresponding to the zeros of the form factor $F_{N,\Delta}(qL)$ of Eq.~\eqref{fgaps}. So we place ourselves in the situation where we have already performed a run in the previous mode and we want to scan further parameter space by a small modification of the set up. 
The dips of $F_{N,\Delta}(qL)$ correspond to the unavoidable dips of the single magnet form factor $F_{\rm single}(q\ell)$, i.e. $q\ell/2=k \pi, k\in \mathbb Z^+$ plus the dips related to the sum of the different magnet contributions, i.e. $q N(\ell+\Delta)/2=k' \pi, k'\in \mathbb Z^+$. 

To get rid of these dips we have four different techniques at our disposal, namely a) introducing phase shift plates, b) a buffer gas in the oscillation region, c) alternating the magnet polarity and d) changing the gap size. 
The first two techniques can be applied to optimize the dips of the single magnet form factor, i.e. can be used to optimize $F_{\rm single}(q\ell)$. The remaining two can never avoid these zeros. 
An optimal technique to cure all these dips in $F_{\rm single}(q\ell)$ with an extra measurement with buffer gas in the oscillation region was described and implemented in~\cite{Ehret:2010mh}.
With phase shift plates a better result is in principle possible but the optical dispersion caused by them (we would need in principle one per magnet) will affect the power build-up of the cavity more than the scattering on the rarefied gas required by the buffer gas technique, so in principle the latter is preferred. 

We are now left with the zeros produced by the interference of the {ALP} waves in the different magnets. 
At low masses, now meaning $q\ell/2< \pi$, i.e. below the first zero of the single magnet form factor we can improve substantially 
by performing our ALP search with the different configurations alternating the magnets. As we have seen, this procedure allows
almost to maximize the form factor (one can get $F\sim 2/\pi$ ) around certain values of $q\ell/2$ in the low mass region, 
cf. for instance Fig.~\ref{fig:examples}. 

We can finally play with the size of the gaps. As we have seen, they allow to shift the pattern of zeros and maxima of the form factor. 
However, we find ourselves between a rock and a hard place.  On the one hand we would like to have as much magnetized region as possible (respecting the minimization of $g_{\rm sens}$), for instance in the above example we would chose $N=12$ instead of $N=10$ LHC dipole magnets. On the other hand, we would also like to have as much gap space to widen as much as possible the sensitivity peaks.
A mixed solution would be to perform a first search with $N=12$ and then change to $N=11$ or $N=10$ without changing the cavity length, such that the range of available gap sizes is bigger.}

\begin{figure}[h!]
\centering
\includegraphics[width=0.45\textwidth]{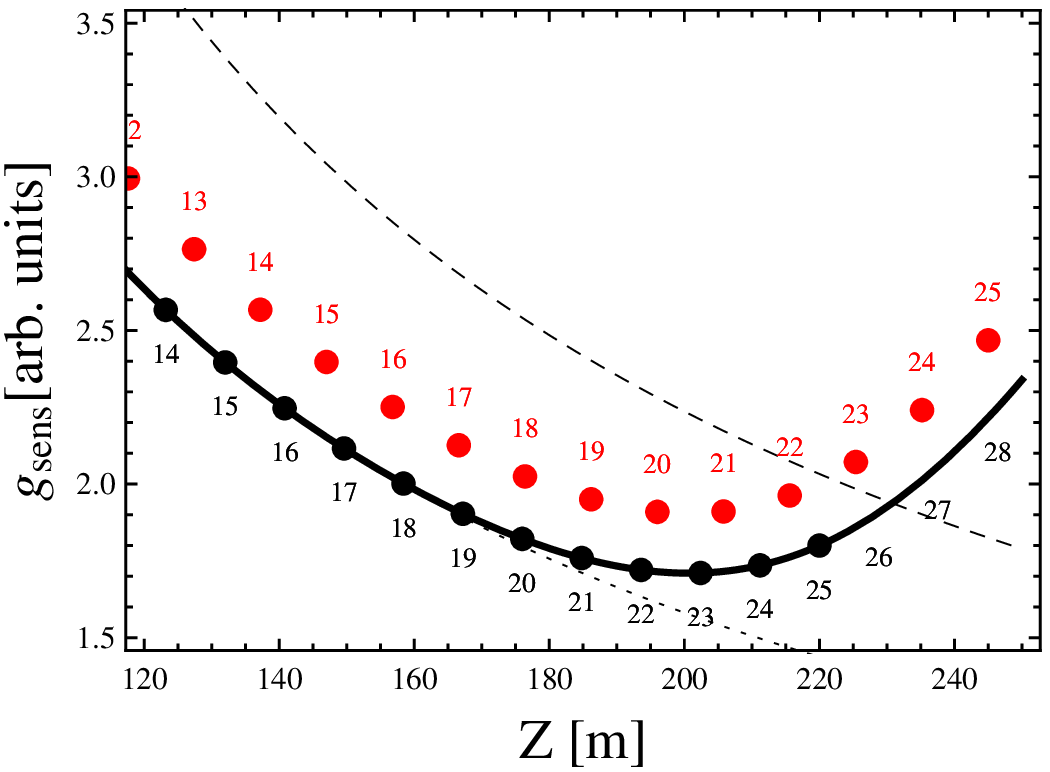}
\includegraphics[width=0.45\textwidth]{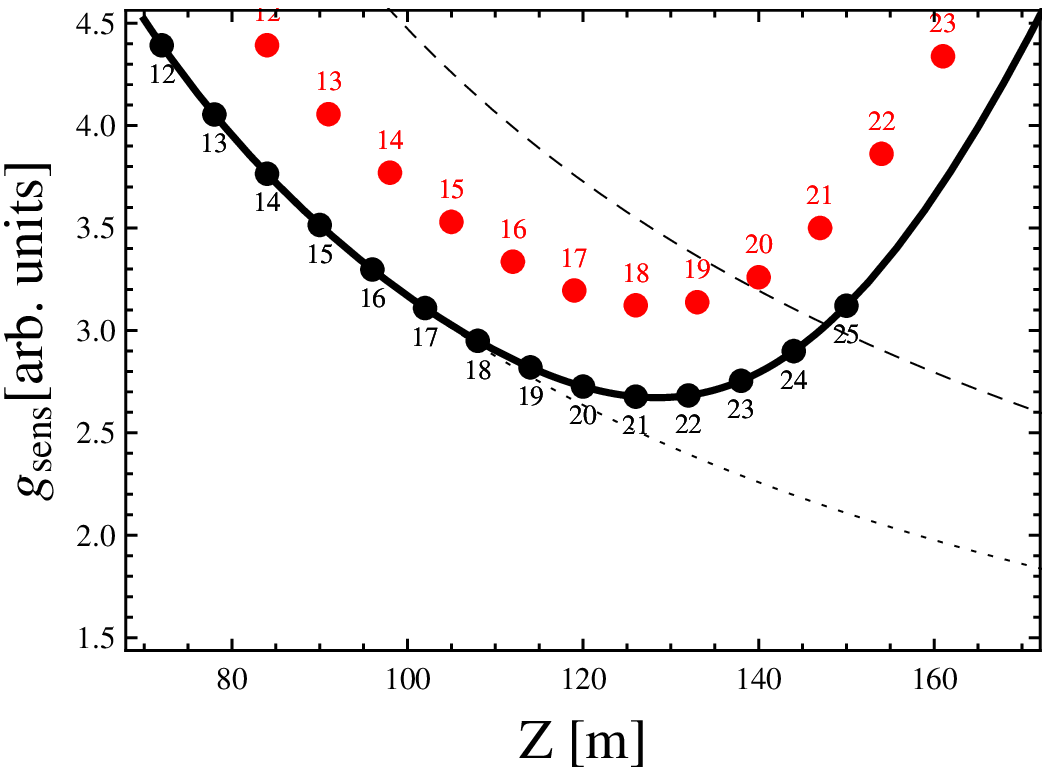}
\caption{\footnotesize{
As before, the solid line shows the optimal case in which the whole cavity length, $Z$, can be filled with magnets $Z=L$.
The black dots indicate the points in which $L$ is an integer multiple of the magnet size, $\ell$.
The red points show the configurations when the gap between the magnets increases to $\Delta=1$ m, ($Z=N(\ell+\Delta)$). 
The dotted line indicates an ideal case in which roundtrip losses would be negligible, $\delta_0=0$ or $a=\infty$. 
The dashed line assumes $\delta_0=\delta_2$.
LEFT: HERA setup. RIGHT: Tevatron setup. 
}
\label{fig:opti2}}
\end{figure}

In Fig.~\ref{fig:opti} (right) we show how the different set ups loose their maximum sensitivity as we increase the gap size. 
When the total length reaches the $160\sim 170$ m figure, the sensitivities start to degrade very much with increasing $Z$
since the clipping losses augment exponentially over the mirror transmissivity losses.
We find for instance that using $N=10$ we can afford gaps up to $\Delta=4$ m and still keep the minimum $g_{\rm sens}$ (which we take here as an indicator of the best attainable sensitivity) comparable to the $N=12$ configuration with a maximum of $\Delta=2$ m gaps.

In summary, a whole {LSW} program could be: find the optimal number of magnets with the requirements on the best attainable power build-up and 
magnet aperture. 
Perform the experiment with all possible wiggler configurations. 
Then, change the gap size to a value that does not spoil the whole sensitivity too much and perform experiments in all the possible configurations again, this would broaden the sensitivity peaks of the different wiggler configurations. The results so far still present the uncomfortable dips caused by the single magnet form factor. 
In order to cope with this, shift all the dips by half an oscillation length by introducing a buffer gas as in~\cite{Ehret:2010mh} and 
perform the experiments above again.

\begin{figure}[t]
\centering
\includegraphics[width=0.5\textwidth]{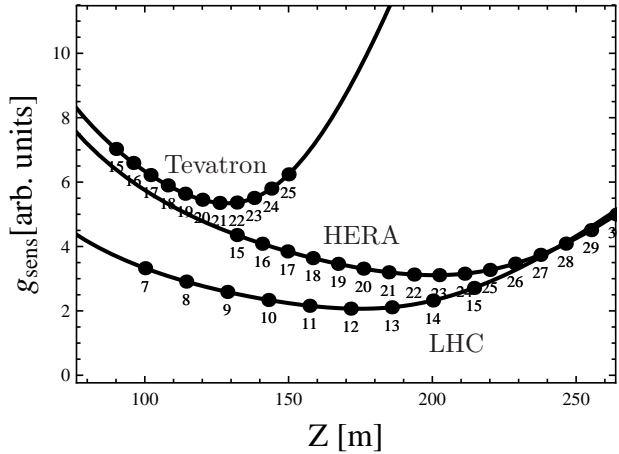}
\Text(-115,83)[l]{ \footnotesize{HERA}}
\Text(-75,42)[l]{ \footnotesize{LHC}}
\Text(-175,110)[l]{ \footnotesize{Tevatron}} 
\caption{\footnotesize{Comparison of $g_{\rm sens}$ vs. $Z$ between three possible LSW setups with HERA, LHC, and Tevatron magnets, in the ideal scenario of no gap in-between the magnets. 
}}
\label{fig:comparison}
\end{figure}

For completeness, we show in Fig.~(\ref{fig:opti2}) a similar optimization analysis for HERA and Tevatron magnets. 
Finally, in Fig.~(\ref{fig:comparison}) we have compared the three types of magnets, in an ideal gapless situation. 
As can be inferred from this last plot, the minimum of $g_{\rm sens}$ can be shifted nearly by one or two magnets, leading to an erroneous estimation of the sensitivity.


{\subsection{Example setups}}
In Table \ref{table:datanext} we show the estimated optimum cavity length for different superconducting dipole magnets currently available, the maximum number of magnets that could fix in the cavity, $N$, and the expected sensitivity in coupling constant, based in our analysis from the previous section.

\begin{figure}[t]
\centering
\includegraphics[width=0.8\textwidth]{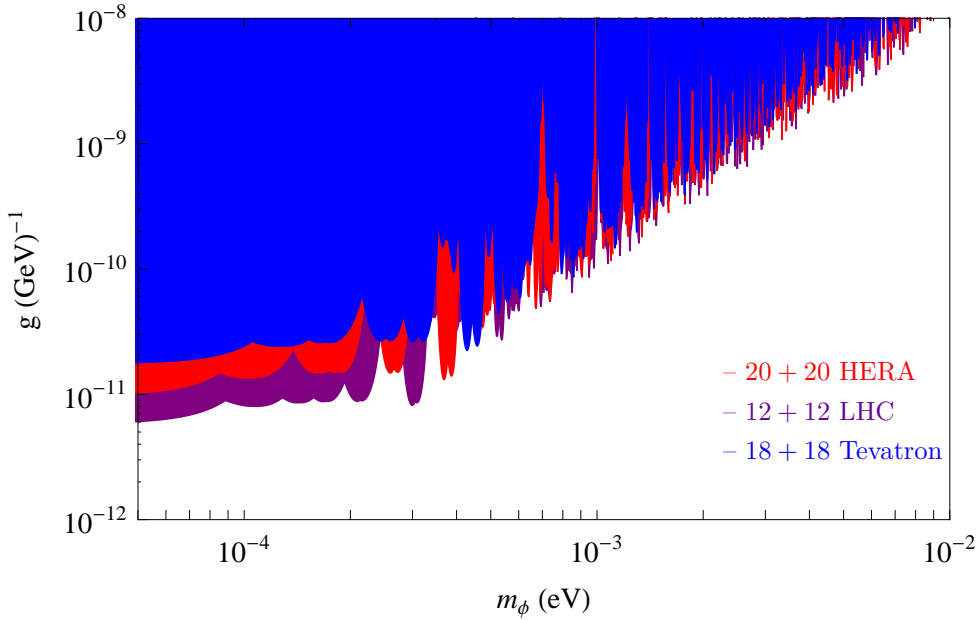}
\Text(-100,95)[l]{ \footnotesize{\color{red}{--~$20+20$ HERA}}}
\Text(-100,80)[l]{ \footnotesize{\color{purple}{--~$12+12$ LHC}}}
\Text(-100,65)[l]{ \footnotesize{\color{blue}{--~$18+18$ Tevatron}}}
\caption{\footnotesize{Estimate of feasible {ALP} bounds for the three types of superconducting dipole magnets currently available}}
\label{fig:comparison2}
\end{figure}

{The LHC magnets lead} with a coupling constant of $5.8\times 10^{-12}$~GeV$^{-1}$, but it is interesting that the results 
obtained {for HERA and Tevatron magnets} are not that far away. Using LHC magnets, and considering a gap between the magnets of 1 m, it would be possible to use {$12+12$ magnets, as discussed before}, without losing the desired $\pb\sim10^{5}$. For HERA and Tevatron it is possible to arrange optimal configurations of {$20+20$ and $18+18$}, respectively. 
In {Fig.}~\ref{fig:comparison2} we summarize this configurations for the three types of superconducting dipole magnets. 
We have assumed power build-up, power incident, etc, as in Tab.~\ref{table:table1}. The different ``wiggler'' configurations have been considered, according to the number of magnets from Table~\ref{table:datanext}.
{
\begin{table}
\centering
\begin{tabular}{|l||c||c||c||c||c|} \hline
Sc-dipole magnet  & $B$ (T)  & $a$ (mm) & $L_{\rm opt}$ (m)& $N$ & $g_{\rm sens}$ (GeV$^{-1}$) \\
\hline
HERA {($\ell=8.8$~m)}& 5.5 &  {30} & {200} & {20} & 9.8 $\times 10^{-12}$ \\
\hline
{Tevatron {($\ell=6$~m)}} & 5 & 24 & {128} & {18} & 1.7~$\times 10^{-11}$ \\
\hline
LHC {($\ell=14.3$~m)} & 9.5  & 28 & {175} & {12} & 5.8~$\times 10^{-12}$\\
\hline
\end{tabular}
\caption{{\footnotesize{Optimal cavity length for three different types of superconducting dipole magnets. The bore aperture radius $a$ given assumes straightened magnets. The bounds on $g_{\rm sens}$ have been computed using the benchmark values given in Table~\ref{table:table1}.}}}
\label{table:datanext}
\end{table}
}
\begin{figure}[t]
\centering \subfigure[]{\label{fig:aplot}\includegraphics[width=0.45\textwidth]{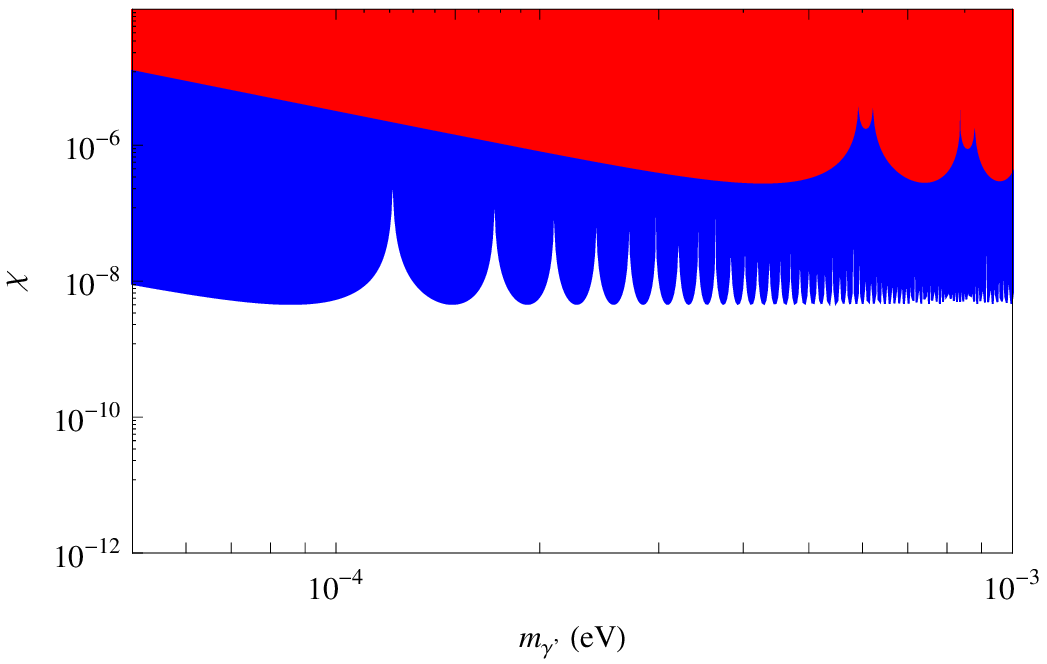}}
\subfigure[]{\label{fig:bplot}\includegraphics[width=0.45\textwidth]{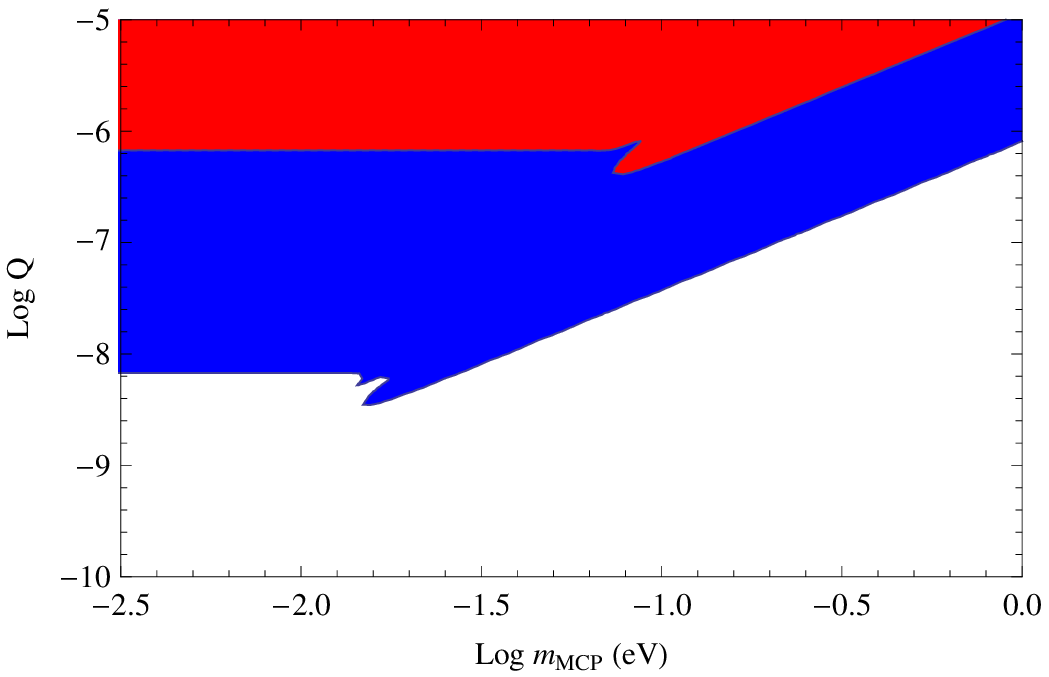}} \caption{\footnotesize{a) Exclusion
plot for massless hidden photons. The red area corresponds to the current bounds from LSW experiments. Blue area
corresponds to the expected exclusion plot for the next generation of LSW. The probability for the last has been computed
according to Eq.~(\ref{regphoton}). b) Exclusion plot for hidden photons $+$ MCPs. The red area corresponds to the latest
bound and the blue area corresponds to the expected sensitivity of the next generation of LSW experiments, according to Table \ref{table:datanext}. We have assumed $e_h=e$. For both plots, we have assumed the optimal cavity of the HERA ($200$~m) and the benchmark input of Table~\ref{table:table1}.  }}
\label{fig:newhiddenmcps}
\end{figure}

In the case of massive hidden photons, the inclusion of the gaps does not affect the sensitivity of the experiment, since the mixing is not affected by the magnetic field. The exclusion plot, considering the {optimal} cavities for the next generation is shown in Fig~\ref{fig:newhiddenmcps}. We have chosen just the HERA configuration for illustration reasons, since all the three configurations are quite similar.

In the case of massless hidden photon and extra mini-charged particles, the separation between the magnets can - in
principle - affect the sensitivity of the experiment. In this case, the computation of a generic formula for $N$ magnets
is more complicated. However, we proceeded adding successive gaps to the all order formula, Eq.~(\ref{allor}). For
instance, considering just one gap, we have found that their contribution appears in the  probability of conversion
$P_{\gamma\rightarrow\gamma'}$, at first order in $\chi$, as
\bb P_{\gamma\rightarrow\gamma'}=\chi^2 \left|1-\mbox{exp}\left( i N~L~\omega \mu^2\right)-\frac{\mu^4}4 f\left(\mu^2,\omega,L,\Delta\right)\right|^2.\ee With $\mu^2= e_h^2 \Delta N_i$, where $\Delta N_i$ accounts for the complex refractive index induced by the creation of a MCP pair and $f$ is a function in the given parameters.

Therefore, for weak mixing - as it is expected - it is possible to show that increasing the number of gaps does not change this kind of suppression. The exclusion plots expected for this type of interaction {is shown} in {Fig.~\ref{fig:newhiddenmcps}}~(b).

\section{Conclusions}\label{conclusions}

The next generation of light shining through walls experiments will most likely use arrays of superconducting dipole magnets 
currently available from the large particle colliders such as HERA, Tevatron or LHC. The engineering of such arrays might require 
the existence of gaps of zero magnetic field in the generation and regeneration regions of the experiment. We have shown 
that the presence of these gaps affects the photon$\leftrightarrow$ALP oscillation probability and how to take advantage of them 
to improve the oscillation probability in certain regions of the ALP parameter space where a gapless configuration would be disastrous. 
This technique is similar to introducing phase shift plates or a buffer gas in the oscillation regions, but benefits from the absence of optical losses, since no dispersive element is introduced in the path of the light.

In order to maximize the sensitivity those experiments will have to {use} optical cavities encompassing each oscillation region. 
The finite aperture of the available magnets {poses constraints on} the transversal extent of the optical modes that can be used in the cavities and this in turn limits the length of the whole experiment by the beam's divergence. {Using magnets with the largest bore aperture is therefore desiderable}.
We have taken all these factors into account and proposed optimized setups for the use of HERA, Tevatron or LHC magnets, cf. Tab.~\ref{table:datanext}.  
The reach of these experiments is shown in Figs.~\ref{fig:comparison2} and~\ref{fig:newhiddenmcps} for ALPs and other WISPs such as hidden photons and minicharged particles. 
In the case of {ALPs, it} constitutes a magnificent improvement upon the currently experimental limits pervading even the current astrophysical limits and reaching a region of parameter space ($g\sim 10^{-11}$ GeV$^{-1}$ at small masses) which has been linked to the resolution of a number of recently raised astrophysical conundrums.
{The realization of such an LSW experiment currently seems to be the most promising purely laboratory probe capable of testing the ultralight ALP hypothesis. Moreover, we believe it can lead the way towards experiments sensitive to a $\sim$meV mass QCD axion.}

\section*{Acknowledgements}

We would like to thank all the members of the ALPS collaboration, especially to Axel Lindner, Tobias Meier and Benno Willke, 
for valuable suggestions and comments.


\begin{thebibliography}{9}

{
\bibitem{Peccei:1977hh}
  R.~D.~Peccei and H.~R.~Quinn,
  Phys.\ Rev.\ Lett.\  {\bf 38}, 1440 (1977).

\bibitem{Weinberg:1977ma}
S.~Weinberg,
   Phys.\ Rev.\ Lett.\ {\bf40}, 223 (1978).

\bibitem{Wilczek:1977pj}
F.~ Wilczek,
 Phys.\ Rev.\ Lett.\  {\bf40}, 279 (1978).

\bibitem{Kim:1979if}
J.~E.~Kim,
  Phys.\ Rev.\ Lett.\ {\bf43}, 103 (1979).

\bibitem{Dine:1981rt}
  M.~Dine, W.~Fischler and M.~Srednicki,
  Phys.\ Lett.\  B {\bf 104}, 199 (1981).

\bibitem{Shifman:1979if}
  M.~A.~Shifman, A.~I.~Vainshtein and V.~I.~Zakharov,
  Nucl.\ Phys.\  B\ {\bf 166}, 493 (1980).

\bibitem{Zhitnitsky:1980tq}
  A.~R.~Zhitnitsky
  Sov.\ J.\ Nucl.\ Phys.\  {\bf 31}, 260 (1980)
  [Yad.\ Fiz.\  {\bf 31}, 497 (1980)].
}

\bibitem{Witten:1984dg}
 E.~Witten,
   Phys.\ Lett.\  B {\bf 149}, 351 (1984).

\bibitem{Conlon:2006tq}
  J.~P.~Conlon,
  JHEP {\bf 0605}, 078 (2006).

\bibitem{Svrcek:2006yi}
 P.~Svrcek and E.~Witten,
  {JHEP} {\bf 0606}, 051 (2006).

\bibitem{Arvanitaki:2009fg}
  A.~Arvanitaki, S.~Dimopoulos, S.~Dubovsky, N.~Kaloper and J.~March-Russell,
  Phys.\ Rev.\  D {\bf 81}, 123530 (2010)
  [arXiv:0905.4720 [hep-th]].

\bibitem{Okun:1982xi}
L.~B.~Okun,
Sov.\ Phys.\ JETP {\bf 56}, 502 (1982)
[Zh.\ Eksp.\ Teor.\ Fiz.\ {\bf 83}, 892 (1982)].

\bibitem{Holdom:1985ag}
B.~Holdom,
Phys.\ Lett.\ B {\bf 166}, 196 (1986).

{
\bibitem{Dienes:1996zr}
 K.~R.~Dienes, C.~F.~Kolda and J.~March-Russell,
  Nucl.\ Phys.\  B {\bf 492}, 104 (1997)
  [arXiv:hep-ph/9610479].

\bibitem{Abel:2003ue}
  S.A.~Abel and B.~W.~Schofield,
  Nucl.\ Phys.\  B {\bf 685}, 150 (2004)
  [arXiv:hep-th/0311051].

\bibitem{Abel:2006qt}
  S.~A.~Abel, J.~Jaeckel, V.~V.~Khoze and A.~Ringwald,
  Phys.\ Lett.\  B {\bf 666}, 66 (2008)
  [arXiv:hep-ph/0608248].

\bibitem{Abel:2008ai}
   S.~A.~Abel, M.~D.~Goodsell, J.~Jaeckel, V.~V.~Khoze and A.~Ringwald,
  JHEP {\bf 0807}, 124 (2008)
  [arXiv:0803.1449 [hep-ph]].

\bibitem{Goodsell:2009xc}
M.~Goodsell, J.~Jaeckel, J.~Redondo and A.~Ringwald,
JHEP {\bf 0911}, 027 (2009)
 [arXiv:0909.0515 [hep-ph]].

\bibitem{Goodsell:2010ie}
  M.~Goodsell and A.~Ringwald,
  Fortsch.\ Phys.\  {\bf 58}, 716 (2010)
  [arXiv:1002.1840 [hep-th]].

}

\bibitem{Sikivie:1983ip}
P.~Sikivie,
Phys.\ Rev.\ Lett.\ {\bf 51}, 1415 (1983)
[Erratum-ibid.\ {\bf 52}, 695 (1984)].

\bibitem{Anselm:1987vj}
A.~A.~Anselm,
Phys.\ Rev.\ D {\bf 37}, 2001 (1988).

\bibitem{VanBibber:1987rq}
K.~Van Bibber {\it et al.}, 
Phys.\ Rev.\ Lett.\ {\bf 59}, 759 (1987).

{
\bibitem{Masso:2005ym}
  E.~Masso and J.~Redondo,
  JCAP {\bf 0509}, 015 (2005)
  [arXiv:hep-ph/0504202].

\bibitem{Jaeckel:2006id}
 J.~Jaeckel, E.~Masso, J.~Redondo, A.~Ringwald and F.~Takahashi,
  arXiv:hep-ph/0605313;
   Phys.\ Rev.\  {\bf D 75}, 013004 (2007)
  [arXiv:hep-ph/0610203].

\bibitem{Masso:2006gc}
  E.~Masso and J.~Redondo,
 Phys.\ Rev.\ Lett.\  {\bf 97}, 151802 (2006)
  [arXiv:hep-ph/0606163].

\bibitem{Redondo:2008tq}
  J.~Redondo
  arXiv:0807.4329 [hep-ph].

\bibitem{Mohapatra:2006pv}
  R.~N.~Mohapatra and S.~Nasri,
  Phys.\ Rev.\ Lett.\  {\bf 98}, 050402 (2007)
  [arXiv:hep-ph/0610068].

\bibitem{Brax:2007ak}
  P.~Brax, C.~van de Bruck and A.~C.~Davis,
   Phys.\ Rev.\ Lett.\  {\bf 99}, 121103 (2007)
  [arXiv:hep-ph/0703243].

\bibitem{Jain:2005nh}
  P.~Jain, S.~Mandal,
   Int.\ J.\ Mod.\ Phys.\  {\bf D} 15, 2095 (2006)
  [arXiv:astro-ph/0512155].

\bibitem{Jain:2006ki}
  P.~Jain and S.~Stokes,
  arXiv:hep-ph/0611006.

\bibitem{Kim:2007wj}
  J.~E.~Kim,
  Phys.\ Rev.\  D {\bf 76}, 051701 (2007)
  [arXiv:0704.3310 [hep-ph]].



\bibitem{Ringwald:2010yr}
  A.~Ringwald,
  arXiv:1003.2339 [hep-ph].

\bibitem{Baker:2010ma}
  O.~K.~Baker, G.~Cantatore, J.~Jaeckel and G.~Mueller,
  arXiv:1007.1835 [Unknown].

\bibitem{Andriamonje:2009dx}
  S.~Andriamonje {\it et al.}  [CAST Collaboration],
  JCAP {\bf 0912}, 002 (2009)
  [arXiv:0906.4488 [hep-ex]].
}

\bibitem{Csaki:2001yk}
C.~Csaki, N.~Kaloper and J.~Terning,
Phys.\ Rev.\ Lett.\ {\bf 88}, 161302 (2002).



\bibitem{Csaki:2003ef}
C.~Csaki, N.~Kaloper, M.~Peloso and J.~Terning,
JCAP {\bf 0305}, 005 (2003).

\bibitem{Mirizzi:2007hr}
A.~Mirizzi, G.~G.~Raffelt and P.~D.~Serpico,
Phys.\ Rev.\ D {\bf 76}, 023001 (2007).

\bibitem{Hooper:2007bq}
D.~Hooper and P.~D.~Serpico,
Phys.\ Rev.\ Lett.\ {\bf 99}, 231102 (2007).
\bibitem{Hochmuth:2007hk}
K.~A.~Hochmuth and G.~Sigl,
Phys.\ Rev.\ D {\bf 76}, 123011 (2007).

\bibitem{Payez:2008pm}
A.~Payez, J.~R.~Cudell and D.~Hutsemekers,
AIP Conf.\ Proc.\ {\bf 1038}, 211 (2008).

\bibitem{Fairbairn:2009zi}
M.~Fairbairn, T.~Rashba and S.~V.~Troitsky,
arXiv:0901.4085 [astro-ph.HE].
\bibitem{Mirizzi:2009nq}
A.~Mirizzi, J.~Redondo and G.~Sigl,
JCAP {\bf 0908}, 001 (2009).
\bibitem{Isern:2008nt}
J.~Isern, E.~Garcia-Berro, S.~Torres and S.~Catalan,
arXiv:0806.2807 [astro-ph].

\bibitem{DeAngelis:2007dy}
A.~De Angelis, O.~Mansutti and M.~Roncadelli,
Phys.\ Rev.\ D {\bf 76}, 121301 (2007).

\bibitem{DeAngelis:2008sk}
A.~De Angelis, O.~Mansutti, M.~Persic and M.~Roncadelli,
arXiv:0807.4246 [astro-ph].

{
\bibitem{Mirizzi:2009aj}
  A.~Mirizzi and D.~Montanino,
  JCAP {\bf 0912}, 004 (2009)
  [arXiv:0911.0015 [astro-ph.HE]].

\bibitem{Bassan:2010ya}
  N.~Bassan, A.~Mirizzi and M.~Roncadelli,
  JCAP {\bf 1005}, 010 (2010)
  [arXiv:1001.5267 [astro-ph.HE]].



\bibitem{Duffy:2006aa}
 L.~D.~Duffy {\it et al.},
Phys.\ Rev.\ D {\bf 74}, 012006 (2006). [arXiv:astro-ph/0603108].

\bibitem{Jaeckel:2010ni}
  J.~Jaeckel and A.~Ringwald,
  arXiv:1002.0329 [hep-ph].

}

{
\bibitem{Ehret:2009sq}
  K.~Ehret {\it et al.}  [ALPS collaboration],
  Nucl.\ Instrum.\ Meth.\  A {\bf 612}, 83 (2009)
  [arXiv:0905.4159 [physics.ins-det]].

\bibitem{Ehret:2010mh}
K.~Ehret {\it et al.},
Phys.\ Lett.\ B {\bf 689} (2010) 149
[arXiv:1004.1313 [Unknown]].
}

\bibitem{Robilliard:2007bq}
  C.~Robilliard, R.~Battesti, M.~Fouche, J.~Mauchain, A.~M.~Sautivet, F.~Amiranoff and C.~Rizzo,
  Phys.\ Rev.\ Lett.\  {\bf 99}, 190403 (2007)
  [arXiv:0707.1296 [hep-ex]].
\bibitem{Fouche:2008jk}
  M.~Fouche {\it et al.},
  Phys.\ Rev.\  D {\bf 78}, 032013 (2008)
  [arXiv:0808.2800 [hep-ex]].

\bibitem{Ruoso:1992nx}
  G.~Ruoso {\it et al.},
  Z.\ Phys.\  C {\bf 56}, 505 (1992).
\bibitem{Cameron:1993mr}
  R.~Cameron {\it et al.},
  Phys.\ Rev.\  D {\bf 47}, 3707 (1993).


\bibitem{Chou:2007zzc}
A.~S.~Chou {\it et al.} [GammeV (T-969) Collaboration],
Phys.\ Rev.\ Lett.\ {\bf 100}, 080402 (2008).

\bibitem{Afanasev:2008fv}
  A.~Afanasev {\it et al.},
  Phys.\ Lett.\  B {\bf 679} (2009) 317
  [arXiv:0810.4189 [hep-ex]].

\bibitem{Afanasev:2008jt}
  A.~Afanasev {\it et al.},
  Phys.\ Rev.\ Lett.\  {\bf 101} (2008) 120401
  [arXiv:0806.2631 [hep-ex]].


\bibitem{Pugnat:2007nu}
P.~Pugnat {\it et al.} [OSQAR Collaboration],
Phys.\ Rev.\ D {\bf 78}, 092003 (2008).



\bibitem{Ringwald:2003nsa}
  A.~Ringwald,
  Phys.\ Lett.\  B {\bf 569}, 51 (2003)
  [arXiv:hep-ph/0306106].



\bibitem{Hoogeveen:1990vq}
F.~Hoogeveen and T.~Ziegenhagen,
Nucl.\ Phys.\ B {\bf 358}, 3 (1991).

\bibitem{Sikivie:2007qm}
P.~Sikivie, D.~B.~Tanner and K.~van Bibber,
Phys.\ Rev.\ Lett.\ {\bf 98}, 172002 (2007).


{
\bibitem{Afanasev:2006cv}
  A.~V.~Afanasev, O.~K.~Baker and K.~W.~McFarlane,
  arXiv:hep-ph/0605250.
{
\bibitem{muellertalk}
G.~Mueller, ``Resonantly Enhanced Axion-Photon Regeneration'', talk presented at Axions 2010, http://www.phys.ufl.edu/research/Axions2010/
}

\bibitem{Raffelt:1987im}
  G.~Raffelt and L.~Stodolsky,
  Phys.\ Rev.\  D {\bf 37}, 1237 (1988).

\bibitem{Adler:2008gk}
S.~L.~Adler, J.~Gamboa, F.~Mendez and J.~Lopez-Sarrion,
Annals Phys.\ {\bf 323}, 2851 (2008)
[arXiv:0801.4739 [hep-ph]].


\bibitem{Ahlers:2007rd}
  M.~Ahlers, H.~Gies, J.~Jaeckel, J.~Redondo and A.~Ringwald,
  Phys.\ Rev.\  D {\bf 76}, 115005 (2007)
  [arXiv:0706.2836 [hep-ph]].

\bibitem{Ahlers:2007qf}
  M.~Ahlers, H.~Gies, J.~Jaeckel, J.~Redondo and A.~Ringwald,
  Phys.\ Rev.\  D {\bf 77}, 095001 (2008)
  [arXiv:0711.4991 [hep-ph]].


\bibitem{Ahlers:2008qc}
  M.~Ahlers, J.~Jaeckel, J.~Redondo and A.~Ringwald,
  Phys.\ Rev.\  D {\bf 78}, 075005 (2008)
  [arXiv:0807.4143 [hep-ph]].


\bibitem{Mueller:2009wt}
 G.~Mueller, P.~Sikivie, D.~B.~Tanner and K.~van Bibber,
Phys.\ Rev.\ D {\bf 80}, 072004 (2009) [arXiv:0907.5387 [hep-ph]].

\bibitem{Bityukov:2000tt}
  S.~I.~Bityukov and N.~V.~Krasnikov,
  Nucl.\ Instrum.\ Meth.\  A {\bf 452} (2000) 518.

\bibitem{Bityukov:1998ju}
  S.~I.~Bityukov and N.~V.~Krasnikov,
  Mod.\ Phys.\ Lett.\  A {\bf 13} (1998) 3235.


}




\bibitem{Jaeckel:2007gk}
  J.~Jaeckel and A.~Ringwald,
  Phys.\ Lett.\  B {\bf 653} (2007) 167
  [arXiv:0706.0693 [hep-ph]].


{




\bibitem{Rabadan:2005dm}
  R.~Rabadan, A.~Ringwald and K.~Sigurdson,
  Phys.\ Rev.\ Lett.\  {\bf 96}, 110407 (2006)
  [arXiv:hep-ph/0511103].

\bibitem{Dias:2009ph}
  A.~G.~Dias and G.~Lugones,
  Phys.\ Lett.\  B {\bf 673}, 101 (2009)
  [arXiv:0902.0749 [hep-ph]].


\bibitem{Battesti:2010dm}
  R.~Battesti {\it et al.},
  arXiv:1008.2672 [hep-ex].
}

\bibitem{siegman}
A.~Siegman, Laser University Science Books, 1986 (Corrected corresponding to errata list published at \mbox{http://standford.edu/\~siegman/lasers\_book\_errata.pdf.})

\bibitem{grav}
H. Luck, et al., Class. Quant. Grav. 23 (2006) 71;
LIGO Collaboration, arXiv:0711.3041v1 [gr-qc].


\end{thebibliography}
\end{document}